\documentclass[letterpaper, preprint, paper,11pt]{AAS}	

\usepackage{bm}
\usepackage{amsmath}
\usepackage{subfigure}
\usepackage[colorlinks=true, pdfstartview=FitV, linkcolor=black, citecolor= black, urlcolor= black]{hyperref}
\usepackage{overcite}
\usepackage{footnpag}			      	
\usepackage[mathscr]{eucal}
\usepackage{bm}
\usepackage{amsfonts}
\definecolor{deepred}{RGB}{210, 0, 0} 

\usepackage{booktabs}
\allowdisplaybreaks

\usepackage{listings}
\usepackage{xcolor}
\lstdefinelanguage{Mathematica}{
  morekeywords={Module, Table, Expand, Coefficient},
  sensitive=true,
  morecomment=[l]{(*},
  morestring=[b]",
}
\PaperNumber{26-899}

\begin{document}

\title{An Efficient Non-Gaussian Chance Constraint Method for Stochastic Nonlinear Problems in Spaceflight}

\author{Ethan R. Burnett\thanks{Assistant Professor, Department of Aerospace Engineering and Engineering Mechanics, University of Texas at Austin, 2617 Wichita St, Austin TX, 78712, USA.}, \ Spencer Boone\thanks{Unaffiliated, Toulouse, FR.},
\ and Niccolò Michelotti\thanks{PhD Student, Dept. of Aerospace Science and Technology, Politecnico di Milano, Via La Masa 34, 20156 Milano IT.}
} 

\maketitle{} 		
\fontsize{10.9pt}{12.72pt}\selectfont

\begin{abstract}
Standard chance-constrained spacecraft guidance typically relies on the assumption that uncertainties in vehicle states obey Gaussian statistics. In frontier applications such as the cislunar environment or deep space flybys, the dynamics can be particularly nonlinear, and time between measurements can be long, leading to the need to make decisions whose outcomes produce non-Gaussian distributions. This paper demonstrates a non-Gaussian confidence boundary technique for stochastic guidance in such applications. Our approach is to consider the true confidence contour as a perturbation of the one predicted from covariance, then to derive perturbed boundary geometry from computed higher-order statistical moments. 
Applying this technique to so-called ``banana-shaped distributions", found in orbital mechanics problems, enables a simple parameterization of the confidence contour using the skew and kurtosis tensors. This parameterization is then applied to a stochastic and nonlinear impulsive spacecraft maneuver targeting problem, with special treatment of a relevant non-convex constraint. 
\end{abstract}

\section{Introduction}
State-of-the-art techniques of guidance and control are often stochastic in nature, whereby control of a nominal trajectory and expected statistical dispersions is jointly enforced. In contrast with many terrestrial robotics applications, spaceflight is often plagued by comparatively large state uncertainties and, sometimes, non-Gaussian statistics. These arise due to operation in dynamic regimes that are nonlinear and chaotic, and long periods without measurements or corrective control maneuvers.

The assumption of Gaussian statistics can be reasonable for some stochastic control problems in spaceflight. Spacecraft rendezvous problems can exploit the nearly linear dynamics of close-proximity relative motion, for which statistical distributions remain very nearly Gaussian. Ref.~\citenum{berning2024BO} leverages this property for passively safe spacecraft rendezvous with a linear covariance (``LinCov") treatment of uncertainty evolution. This enables a ``chance-constrained" approach, whereby satisfaction of the path constraint can be certified to a certain probability level. Chance-constrained approaches have become popular for safety-critical problems in spaceflight with sufficiently small and Gaussian dispersions~\cite{OguriChanceConstr}. The prospect of increasingly autonomous operation in space foresees more frequent measurements and more active maneuvering, thus such approaches may often be satisfactory. However, in long time-horizon maneuver planning, or in cases where accurate measurements are not available for long periods, the specter of non-Gaussian statistics cannot be ignored. 

In spaceflight, a commonly observed manifestation of non-Gaussian statistics is with the so-called ``banana-shaped" distributions. These emerge because the nonlinear dynamics of orbital mechanics tend to first stretch and then bend the ellipsoidal confidence region corresponding to an initially compact (and perhaps also Gaussian) distribution. For Keplerian problems, this problem can be partially avoided via use of ``less nonlinear" coordinates such as polar coordinates or orbit elements (see e.g. Ref.~\citenum{junkAdven}). However, practical path constraints may not be convenient to express in such coordinates, and also non-Keplerian contexts such as cislunar astrodynamics challenge this solution strategy, as superior native coordinates are typically not available nor easy to identify.

For stochastic guidance and control with non-Gaussian statistics in astrodynamics, Monte Carlo methods are often the default strategy, but they are very slow. Overall the field is still in pursuit of accurate and efficient methods for such problems. One strategy is to probe the statistical moments beyond mean and covariance for geometric insights. Methods of estimating these moments include polynomial chaos expansion (PCE)\cite{JonesPCE}, or the conjugate unscented transform (CUT)\cite{CUT_ACC}, which uses comparatively far fewer points and has the benefit of being deterministic. Recently, Ref.~\citenum{qi2025steeringnonlinear} explores nonlinear steering of non-Gaussian distribution back to an approximately Gaussian nature by leveraging feedback on the sigma points for the conjugate unscented transform. This extends the more familiar ``covariance steering" (e.g., Ref.~\citenum{sial2025CovSteering}) to a regulation of higher statistical moments. 

Active statistical steering methods introduce an additional continuous control feedback term which increases fuel use and might not always be operationally feasible. Instead, long-time horizon maneuvers can be planned, accepting the resulting non-Gaussian statistical distribution, and simply enforcing chance constraints on the non-Gaussian distribution itself. Along this line of thought, Ref.~\citenum{boone_cdc} applies a chance constraint approach with Gaussian Mixture Models to render the non-Gaussian distribution, however the approach is still rather numerically intensive. Ref.~\citenum{BurnettBooneCDC26} introduces a promising new technique for analytically approximating confidence boundaries for the non-Gaussian ``banana distributions" encountered in astrodynamics. Namely, select components of the higher-order moments of skew and kurtosis are used to directly inform geometric corrections to the ellipsoidal boundary predicted from covariance in the Gaussian limit. The methodology works well for high-fidelity cases we've tested\cite{AAS26b}, and when paired with an efficient methodology for estimating skew and kurtosis, such as CUT4\cite{CUT_ACC}, is quite numerically efficient. In this paper, which is devoted to application of the method, we focus on planar (two-state) parameterizations, but note that full 6-state distributions can be addressed by linear composition of each pair\cite{{AAS26b}}. 

In this paper, we demonstrate how possession of an accurate analytic approximation for a non-Gaussian confidence boundary allows for explicit consideration of the resulting non-convex constraint. We design a stochastic guidance approach that is fast and stable for admissible non-Gaussian cases with banana-shaped distributions, without excessive conservatism. After introducing the stochastic guidance approach, this paper revisits the numerical example of Ref.~\citenum{BurnettBooneCDC26}, then furnishes all developments in a timely and relevant end-to-end example: long-horizon re-entry corridor targeting of an Artemis II-like lunar free-return trajectory in the multibody Earth-Moon system. 

\section{Derivation of Non-Gaussian confidence contour}
We derive corrections of the confidence bounds in the case of weakly non-Gaussian statistics (i.e. statistics for which the distribution is ``banana-shaped" -- see e.g. Ref.~\citenum{junkAdven}), leveraging these quantities. The approach we will expand on was first outlined in Ref.~\citenum{BurnettBooneCDC26}. 
\subsection{Confidence Bounds: The Gaussian Case, and a Case for Small Corrections}
The covariance ellipse can be used to bound Gaussian distributions with a certain $k\sigma$ confidence level with covariance \(P \in \mathbb{R}^{6\times6}\). A 2$\times$2 subset covariance is extracted as
\begin{equation}
    \Sigma = 
        \begin{bmatrix}
            \Sigma_{xx} & \Sigma_{xy} \\
            \Sigma_{xy} & \Sigma_{yy}
        \end{bmatrix}
    = 
        S\,P\,S^\top,
    \label{eq:sigma_def}
\end{equation}
where $S$ is a selection matrix composed of ones and zeros. Let \(\bm{\mu} = (\bar{x},\,\bar{y})\) denote the mean position.
The confidence ellipse is the set of points \(\bm{r} = (x,y)\) satisfying
\begin{equation}
    (\bm{r}-\bm{\mu})^\top \Sigma^{-1} (\bm{r}-\bm{\mu}) = k^2 .
    \label{eq:ellipse_mahalanobis}
\end{equation}
For a ``3-sigma'' scaled ellipse, \(k=3\).
For a 2D confidence level \(p\), one may use
\(k^2 = {\bm{\chi}}^2_{2,p}\).
The covariance is diagonalized as below:
\begin{subequations}
\begin{align}
    \Sigma = & \ R \Lambda R^\top ,
    \label{eq:Sigma_eigendecomp}
    \\
    \Lambda = & \ \operatorname{diag}(\lambda_1,\,\lambda_2),\qquad
    \lambda_1 \ge \lambda_2 > 0 ,
    \\
    R = & \
        \begin{bmatrix}
            \cos\theta & -\sin\theta \\
            \sin\theta & \cos\theta
        \end{bmatrix}.
\end{align}
\label{eq:R_def}
\end{subequations}
The rotation angle \(\theta\) may be written as $\theta = \frac{1}{2}\operatorname{atan2}
    \!\big( 2\,\Sigma_{xy},\, \Sigma_{xx}-\Sigma_{yy} \big)$. 
The boundary of a certain statistical confidence level is parameterized by an ellipse with the following principal semi-axes:
\begin{equation}
    \overline{a} = k\sqrt{\lambda_1}, \qquad 
    \overline{b} = k\sqrt{\lambda_2} .
    \label{eq:semi_axes}
\end{equation}
We define local principal coordinates $\bm{z} = (u,v)^{\top}$, which point along the long and short axes:
\begin{equation}
    u = \overline{a}\cos t, \qquad v = \overline{b}\sin t , \qquad t\in[0,2\pi) .
    \label{eq:local_uv}
\end{equation}
Thus a point on the elliptical boundary is parameterized as below:
\begin{equation}
    \bm{r}(t) = \bm{\mu} + R
        \begin{pmatrix}
            u(t) \\ v(t)
        \end{pmatrix}
    \label{eq:r_param}
\end{equation}
We additionally define the normalized coordinates $\hat{u} = u/\sqrt{\lambda_{1}}$ and $\hat{v} = v/\sqrt{\lambda_{2}}$, which in the case of the parameterization of a Gaussian boundary satisfy $\hat{u} = k\cos{t}$, $\hat{v} = k\sin{t}$.

We seek methods suitable for when a distribution is ``weakly" non-Gaussian, 
i.e. confidence contours close to the mean still bear strong similarity to those expected from a Gaussian distribution, but further out they lose their accuracy. We seek suitable and tractable first-order corrections (i.e. linear in a parameter). 
The dominant effect commonly observed in astrodynamics problems is the stretching of the distribution along the so-called ``maximum stretching direction", which corresponds with the maximum eigendirection of the Cauchy-Green tensor (e.g., Ref.~\citenum{Boone_STTorbits}).
Beyond linear effects, the first and most noteworthy key feature is the bend of the distribution: the decoupling of $\hat{u}, \hat{v}$ from the Gaussian case is lost. In particular, sufficiently large departures in $\hat{u}$ from the mean will carry a bend in $\hat{v}$. As a first-order correction of the elliptical assumption, we thus apply the following \emph{ansatz} which breaks the independence of the principal coordinates for large deviations:
\begin{equation}
\label{myansatz}
\hat{v}_{q} \approx \beta_{q} + \alpha_{q}\hat{u}_{q}^{2}
\end{equation}
with index $q$ for each possible two-state slice, and parameters $(\alpha_{q},\beta_{q})$ are to be determined. The motivation for the quadratic ansatz is as follows: Viewing the action of the flow of the nonlinear dynamics on the statistical distribution (and hence its confidence boundary) as a weakly nonlinear map, the first expected contribution after the linear effect should be quadratic. 

The next feature is less obvious. The probability density towards $\hat{u}  = +g$ and $\hat{u} = -g$ is no longer even for some $g\gg 0$. In other words, the symmetry of the confidence bounds is broken. This effect is more noticeable along the long axis of the distribution, so we seek a correction to $\hat{u}$ but not $\hat{v}$. This is similar to the logic by which we prioritized bending of the form $\hat{v} \propto \hat{u}^{2}$ but ignored the (assumed sub-dominant) analogous bending term $\hat{u} \propto \hat{v}^{2}$. Focusing just on corrections along $\hat{u}$, the Cornish-Fisher expansion provides an asymptotic approximation of the quantiles of a univariate non-Gaussian distribution based on its cumulants.\cite{HMF} Below we provide the first-order term: 
\begin{equation}
\label{CFish1}
x(p) \approx \mu + \sigma \left(z + \frac{\gamma_{1}}{6}(z^{2}-1) \right)
\end{equation}
where $x$ follows a slightly non-Gaussian univariate distribution, with mean $\mu$ and standard deviation $\sigma$. Furthermore $z=\Phi^{-1}(p)$ where $\Phi$ is the cumulative distribution function of the standard normal distribution, i.e. $\Phi(3)\approx0.9987$. Lastly, $\gamma_{1}$ is expressed in terms of skew and standard deviation:
\begin{equation}
\label{CFish2a}
\gamma_{1} = \frac{\mu_{3}}{\sigma^{3}}, \ \ \ \mu_{n} = \mathbb{E}[(x-\mu)^{n}]
\end{equation}
We seek an additive $t$-periodic correction to the (whitened) Gaussian parameterization $\hat{u}(t) = k\cos{t}$ which obeys the following properties for $c(k)=\frac{\gamma_{1}}{6}(k^{2}-1)$:
\begin{subequations}
\label{CFish2b}
\begin{align}
\delta\hat{u}(0) = \delta\hat{u}(\pi) = & \ c(k) \\
\delta\hat{u}(\frac{\pi}{2}) = \delta\hat{u}(\frac{3\pi}{2}) = & \ 0 \\
\delta\hat{u}'(0) = \delta\hat{u}'(\pi) = & \ 0 \\
\delta\hat{u}'(\frac{\pi}{2}) = \delta\hat{u}'(\frac{3\pi}{2}) = & \ 0 
\end{align}
\end{subequations}
The first two conditions state that along the direction $\hat{u}$, we expect to recover the univariate correction: The $k\sigma$ boundaries shift to $-k + c(k)$ for $\hat{u}<0$ and $k+c(k)$ for $\hat{u}>0$. The second two enforce no net change when $\hat{u} = 0$. 
The last four conditions are for regularity -- to avoid unphysical directional biases. Noting that the univariate Cornish-Fisher expansion tells us nothing about how the confidence contour changes except purely along the long axis of the distribution, we furthermore require evenness of $\delta\hat{u}(t)$ about $t=0$ and $t=\pi$, as any other choice induces off-axis asymmetries that cannot be justified. The class of all functions satisfying these conditions can be expressed as:
\begin{equation}
\label{duCS}
\delta\hat{u}_{q}(t) = \eta_{0,q} + \sum_{n\geq1}\eta_{n,q}\cos{(nt)}
\end{equation}
We will start with the derivation of the proper coefficients for Eq.~\eqref{myansatz}, then revisit this. 

\subsection{Moment Identities}
Let $\bm{X}=(X_1,\dots,X_d)^{\!\top}$ be a random vector with finite central moments through order four, mean $\bm{\mu}$, covariance $\Sigma$, third-order moment tensor $M^{(3)}$, and fourth-order moment tensor $M^{(4)}$. The tensors $\Sigma_{ij}$, $M^{(3)}_{ijk}$, and $M^{(4)}_{ijkl}$ are invariant under permutation of their indices. For a multivariate Gaussian distribution,
\begin{equation}
    M^{(3)}_{ijk}=0,
    \label{eq:M3_gaussian_zero}
\end{equation}
and the fourth central moment is generated by the covariance as
\begin{equation}
    M^{(4)}_{ijkl}
    =
    \Sigma_{ij}\Sigma_{kl}
    +
    \Sigma_{ik}\Sigma_{jl}
    +
    \Sigma_{il}\Sigma_{jk}.
    \label{eq:M4_gaussian_identity}
\end{equation}
Equivalently, the fourth-order cumulant tensor vanishes for Gaussian statistics:
\begin{equation}
    \kappa_{ijkl}
    \equiv
    M^{(4)}_{ijkl}
    -
    \left(
        \Sigma_{ij}\Sigma_{kl}
        +
        \Sigma_{ik}\Sigma_{jl}
        +
        \Sigma_{il}\Sigma_{jk}
    \right)
    \label{eq:kappa_def}
\end{equation}
\begin{equation}
    \kappa_{ijkl}=0
    \qquad
    \forall\, i,j,k,l.
    \label{eq:kappa_gaussian_zero}
\end{equation}
Thus, deviations from Eqs.~\eqref{eq:M3_gaussian_zero} and \eqref{eq:kappa_gaussian_zero} provide third- and fourth-order indicators of non-Gaussianity.

\subsection{First Correction: The bend of the banana}
Referring the reader again to Eq.~\eqref{myansatz}, for a given slice $q$, we solve for parameters $(\alpha_{q}, \beta_{q})$ minimizing the expected square of the fit error below: 
\begin{equation}
(\alpha,\beta)
=
\arg\min_{\alpha,\beta}
\;\mathbb{E}\!\left[\bigl(\hat v - \beta - \alpha\,\hat u^2\bigr)^2\right].
\label{eq:pop_min}
\end{equation}

Applying the first-order conditions of optimality w.r.t. $\alpha$ and $\beta$, we obtain:
\begin{subequations}
\label{eq:pop_normal}
\begin{align}
\mathbb{E}[\hat v] 
&= \beta + \alpha\,\mathbb{E}[\hat u^2],
\label{eq:pop_normal_1}
\\
\mathbb{E}[\hat v\,\hat u^2] 
&= \beta\,\mathbb{E}[\hat u^2] + \alpha\,\mathbb{E}[\hat u^4].
\label{eq:pop_normal_2}
\end{align}
\end{subequations}
These are linear equations in $\alpha,\beta$. Some further manipulations isolate the corrective coefficients: 
\begin{subequations}
\label{eq:pop_solution}
\begin{align}
\alpha 
&= 
\frac{
\mathbb{E}[\hat v\,\hat u^2] - \mathbb{E}[\hat u^2]\,\mathbb{E}[\hat v]
}{
\mathbb{E}[\hat u^4] - \mathbb{E}[\hat u^2]^2
},
\label{eq:pop_alpha}
\\[6pt]
\beta
&=
\frac{
\mathbb{E}[\hat u^4]\,\mathbb{E}[\hat v] - \mathbb{E}[\hat u^2]\,\mathbb{E}[\hat v\,\hat u^2]
}{
\mathbb{E}[\hat u^4] - \mathbb{E}[\hat u^2]^2
}.
\label{eq:pop_beta}
\end{align}
\end{subequations}
For a 2-state sample $\bm{r}$ (not necessarily from a Gaussian distribution) and a given slice covariance $\Sigma_{q}$ of the form of Eq.~\eqref{eq:sigma_def}, we define the following convenient whitening transformation:
\begin{equation}
W_{q}
=
\Lambda_{q}^{-1/2}\,R_{q}^{\top},
\label{eq:W_q}
\end{equation}
\begin{equation}
\begin{pmatrix}
\hat u\\[4pt]
\hat v
\end{pmatrix}
=
W_{q}\bigl(\mathbf r - \boldsymbol{\mu}_{q}\bigr).
\label{eq:uv_hat}
\end{equation}
We seek to apply the transformation of Eq.~\eqref{eq:uv_hat} to the random state from our non-Gaussian distribution, substitute the resulting identities into Eqs.~\eqref{eq:pop_alpha}-\eqref{eq:pop_beta}, and solve for the coefficients $\alpha,\beta$. Further simplification is possible. First it is easy to show that $\mathbb{E}[\hat{u}] = \mathbb{E}[\hat{v}]=0$, leveraging some useful quantities:
\begin{equation}
\label{eq:EfA1}
\bm{a} = W_{q}^{\top}\begin{pmatrix} 1 \\ 0 \end{pmatrix}; 
\bm{b} = W_{q}^{\top}\begin{pmatrix} 0 \\ 1 \end{pmatrix};
\bm{\delta} = \bm{r} - \bm{\mu}_{q}.
\end{equation}
For nonlinear terms we switch to an index notation for convenience: 
\begin{subequations}
\begin{align}
\label{eq:EfA3}
\hat{u}^{2} = & \ \left(\bm{a}^{\top}\bm{\delta}\right)^{2} = \left(\sum_{j}a_{j}\delta_{j}\right)\left(\sum_{k}a_{k}\delta_{k}\right) = \sum_{j}\sum_{k}a_{j}a_{k}\delta_{j}\delta_{k} \\
\hat{v}\hat{u}^{2} = & \ \sum_{i}b_{i}\delta_{i}\cdot\hat{u}^{2} = \sum_{i}\sum_{j}\sum_{k}b_{i}a_{j}a_{k}\delta_{i}\delta_{j}\delta_{k} 
\end{align}
\end{subequations}
From such expressions it is simple to compute the necessary expected values, e.g.:
\begin{equation}
\label{eq:EfA4}
\mathbb{E}[\hat{v}\hat{u}^{2}] = \sum_{i}\sum_{j}\sum_{k}b_{i}a_{j}a_{k}\mathbb{E}[\delta_{i}\delta_{j}\delta_{k}] = \sum_{i,j,k} b_{i}a_{j}a_{k}M^{(3)}_{q,ijk}
\end{equation}
Lastly, we establish one more identity by noting $\Sigma_{q} = R_{q}\Lambda_{q}R_{q}^{\top}$:
\begin{equation}
\label{eq:EfA4b}
\mathbb{E}[\hat{u}^{2}] = \bm{a}^{\top}\Sigma_{q}\bm{a} =  \begin{bmatrix}1 & 0\end{bmatrix}\Lambda_{q}^{-1/2}R_{q}^{\top}\Sigma_{q}R_{q}\left(\Lambda^{-1/2}\right)^{\top}\begin{bmatrix} 1 \\ 0 \end{bmatrix} = 1
\end{equation}
The final result for Eqs.~\eqref{eq:pop_alpha}-\eqref{eq:pop_beta} is thus obtained, solely as a function of skew and kurtosis:
\begin{subequations}
\label{eq:pop_solution2}
\begin{align}
\alpha 
&= \frac{\sum_{i,j,k}b_{i}a_{j}a_{k}M^{(3)}_{q,ijk}}{\sum_{i,j,k,l}a_{i}a_{j}a_{k}a_{l}M_{q,ijkl}^{(4)} - 1},
\label{eq:pop_alpha2}
\\[6pt]
\beta
&= -\alpha.
\label{eq:pop_beta2}
\end{align}
\end{subequations}
These satisfy the expected property $\alpha=\beta=0$ for a purely Gaussian distribution. Finally, we can establish the first corrected parameterization of the confidence contour. Revisiting Eqs.~\eqref{eq:local_uv} and ~\eqref{myansatz}, we leverage the zero-order solution $\hat{u} = k\cos{t}$, and expect that $\hat{v}$ should be unmodified from the Gaussian case in the event that $\alpha=0$ (in other words, we retain an unbiased reference value). Thus:
\begin{equation}
v = \overline{b}\sin{t} + \alpha\sqrt{\lambda_{2}}\left(k^{2}\cos^{2}{t} - 1\right), \qquad t\in[0,2\pi)
    \label{eq:local_uv2}
\end{equation}
where $k$ is the sigma confidence level and the definitions $\overline{a} = k\sqrt{\lambda_1}$ and $\overline{b} = k\sqrt{\lambda_2}$ are reused from earlier and these are unrelated to the components of $\bm{a}$, $\bm{b}$. This result notably recovers the expected ``bending" form, via the cosine-squared term in the $v$ coordinate of the contour. 
\subsection{Second Correction: Long axis asymmetry}
As the first four moments of a non-Gaussian distribution are not enough to uniquely determine the isoprobability contour, we seek the simplest justifiable correction based on known information and constraints. Returning to Eq.~\eqref{duCS}, we establish the simplest non-trivial satisfier of Eq.~\eqref{CFish2b}:
\begin{equation}
\label{myansatzB}
\delta\hat{u} = \eta_{0} + \eta_{2}\cos{2t}
\end{equation}
Applying the constraints, we obtain:
\begin{subequations}
\label{LAcorr1}
\begin{align}
\eta_{0} = & \ \eta_{2} \\
2\eta_{0} = & \ c(k) \\
c(k) = & \ \frac{1}{6}\mathbb{E}[\hat{u}^{3}](k^{2}-1)
\end{align}
\end{subequations}
Or, rewritten below in terms of the skew directly:
\begin{equation}
\label{ckSkew}
c(k)
= \frac{k^2 - 1}{6}
  \sum_{i,j,k}
  a_i\,a_j\,a_k\,M^{(3)}_{q,ijk}
\end{equation}
Thus, applying the double-angle identity and rescaling as $u(t) = \sqrt{\lambda_{1}}\hat{u}(t)$, the final corrected $u(t)$ is:
\begin{equation}
u(t) = \overline{a}\cos{t} + c(k)\sqrt{\lambda_{1}}\cos^{2}{t}
\label{eq:local_uv3}
\end{equation}
As before, this result recovers the Gaussian case if $c(k) = 0$ (no skew). The final corrected contour is obtained by combining Eqs.~\eqref{eq:local_uv2} and \eqref{eq:local_uv3} and mapping back to $\bm{X}$ coordinates via Eq.~\eqref{eq:r_param}. 

\noindent\textbf{\underline{Key Result 1}:}
In our non-Gaussian setting, the ellipse is replaced by the banana contour
\begin{subequations}\label{eq:banana_param_halfplane}
\begin{align}
u(t) &= \overline{a}\cos t + c(k)\sqrt{\lambda_1}\cos^2 t, \label{eq:banana_param_halfplane_u}\\
v(t) &= \overline{b}\sin t + \alpha \sqrt{\lambda_2}\big(k^2\cos^2 t - 1\big), \label{eq:banana_param_halfplane_v}
\end{align}
\end{subequations}
with mapping
\begin{equation}
\bm{r}(t)=\bm{\mu} + R
\begin{pmatrix}
u(t)\\[2pt]
v(t)
\end{pmatrix}.
\label{eq:r_halfplane}
\end{equation}
Thus the confidence contour of the banana distribution is treated as a first-order deformation of the Gaussian linear covariance ellipse. Despite its simplicity, numerical experiments have shown it works well for even very large perturbed distributions in astrodynamics problems. As a first-order correction, it can be applied simultaneously to multiple distinct coordinate pairs (e.g. $u$-$v$ and $u$-$w$ for principal directions $u,v,w$). This is discussed in Ref.~\citenum{AAS26b}, along with high-fidelity test cases.
\section{Geometric Constraints}
Now we introduce the local half-plane constraint
\begin{equation}
\bm{n}^\top \bm{r} - b_0 \le 0,
\label{eq:halfplane_constraint}
\end{equation}
where $\bm{n} \in \mathbb{R}^2$ is the outward normal of the local boundary approximation and $b_0 \in \mathbb{R}$ is the associated offset. Define the contour residual
\begin{equation}
\psi({\bm{\chi}},t) \equiv \bm{n}^\top \bm{r}({\bm{\chi}},t) - b_0,
\label{eq:psi_def_halfplane}
\end{equation}
where ${\bm{\chi}}$ denotes the free optimization variables (e.g. initial state/parameter, delta-V, or other control variables). The requirement that the confidence contour lie on the safe side of the half-plane is
\begin{equation}
\max_{t\in[0,2\pi)} \psi({\bm{\chi}},t) \le 0.
\label{eq:support_constraint_halfplane}
\end{equation}
The basic problem geometry is depicted in Fig.~\ref{fig:badorient}(a).

\subsection{Gaussian case}
We first recover the familiar Gaussian result. Let 
$\bm{m}({\bm{\chi}}) \equiv R({\bm{\chi}})^\top \bm{n} =
\begin{pmatrix}
m_1({\bm{\chi}})\\
m_2({\bm{\chi}})
\end{pmatrix}$.
Then, for the Gaussian contour,
\begin{equation}
\psi_G({\bm{\chi}},t)
=
\bm{n}^\top \mu({\bm{\chi}})-b_0
+
{\bm{n}^{\top}} R
\begin{pmatrix}
u_G(t)\\[2pt]
v_G(t)
\end{pmatrix}.
\label{eq:psi_gaussian_expand}
\end{equation}
Applying the definitions of $\bm{m}(\bm{\chi})$, and of $\overline{a}$ and $\overline{b}$ from Eq.~\eqref{eq:semi_axes},
\begin{equation}
\psi_G({\bm{\chi}},t)
=
{\bm{n}^{\top}} \mu({\bm{\chi}})-b_0
+
k\,m_1({\bm{\chi}})\sqrt{\lambda_1({\bm{\chi}})}\cos t
+
k\,m_2({\bm{\chi}})\sqrt{\lambda_2({\bm{\chi}})}\sin t.
\label{eq:psi_gaussian_expand2}
\end{equation}
For any scalars $p,q$, 
$\max_t \big(p\cos t + q\sin t\big)=\sqrt{p^2+q^2}$, so
\begin{align}
\max_t \psi_G({\bm{\chi}},t)
&=
{\bm{n}^{\top}} \mu({\bm{\chi}})-b_0
+
k\sqrt{m_1({\bm{\chi}})^2\lambda_1({\bm{\chi}})+m_2({\bm{\chi}})^2\lambda_2({\bm{\chi}})}.
\label{eq:psi_gaussian_support1}
\end{align}
Furthermore, because $\bm{m}=R^\top \bm{n}$ and $\Sigma = R\Lambda R^{\top}$, it is easy to show $m_1^2\lambda_1+m_2^2\lambda_2 = {\bm{n}^{\top}} \Sigma \bm{n}$. 
Therefore, the Gaussian half-plane chance surrogate is recovered as:
\begin{equation}
\max_t \psi_G({\bm{\chi}},t)
=
{\bm{n}^{\top}} \mu({\bm{\chi}})-b_0
+
k\sqrt{{\bm{n}^{\top}} \Sigma({\bm{\chi}})\,\bm{n}} \leq 0.
\label{eq:gaussian_halfplane_final}
\end{equation}
This convenient geometric constraint facilitates covariance-based strategies for stochastic control.\cite{OguriMcMahoIEEE2019, boone_cdc} 

\subsection{Non-Gaussian case}
Substituting the banana parameterization \eqref{eq:banana_param_halfplane} into \eqref{eq:psi_def_halfplane} gives
\begin{align}
\psi({\bm{\chi}},t)
&=
{\bm{n}^{\top}} \mu({\bm{\chi}})-b_0
+
m_1({\bm{\chi}})\,u(t)
+
m_2({\bm{\chi}})\,v(t) \nonumber\\
&=
{\bm{n}^{\top}} \mu({\bm{\chi}})-b_0
+
m_1\Big(a\cos t + c(k,{\bm{\chi}})\sqrt{\lambda_1}\cos^2 t\Big)
\\ & \ \ \ \ \ \ \ \ \ \ \ \ \ \ \ \ \ + m_2\Big(b\sin t+\alpha({\bm{\chi}})\sqrt{\lambda_2}(k^2\cos^2 t-1)\Big).
\label{eq:psi_banana_expand}
\end{align}
Grouping terms by trigonometric dependence yields
\begin{equation}
\psi({\bm{\chi}},t)
=
A({\bm{\chi}})+B({\bm{\chi}})\cos t+C({\bm{\chi}})\sin t+D({\bm{\chi}})\cos^2 t,
\label{eq:psi_ABCD}
\end{equation}
with coefficients defined as below:
\begin{subequations}\label{eq:ABCD_defs}
\begin{align}
A({\bm{\chi}})&={\bm{n}^{\top}} \mu({\bm{\chi}})-b_0-m_2({\bm{\chi}})\,\alpha({\bm{\chi}})\sqrt{\lambda_2({\bm{\chi}})}, \label{eq:A_def}\\
B({\bm{\chi}})&=k\,m_1({\bm{\chi}})\sqrt{\lambda_1({\bm{\chi}})}, \label{eq:B_def}\\
C({\bm{\chi}})&=k\,m_2({\bm{\chi}})\sqrt{\lambda_2({\bm{\chi}})}, \label{eq:C_def}\\
D({\bm{\chi}})&=m_1({\bm{\chi}})c(k,{\bm{\chi}})\sqrt{\lambda_1({\bm{\chi}})}
+
m_2({\bm{\chi}})\alpha({\bm{\chi}})k^2\sqrt{\lambda_2({\bm{\chi}})}. \label{eq:D_def}
\end{align}
\end{subequations}
Note that the banana correction contributes only through the constant and the $\cos^2 t$ terms. With the addition of this term, we find no way to recover an analog for Eq.~\eqref{eq:psi_gaussian_support1}, blocking recovery of a non-Gaussian analog for Eq.~\eqref{eq:gaussian_halfplane_final}. This is due to branching behavior in the angle parameter $t$ of the contour yielding the worst-case value of the half plane constraint. 

Two major points are worth emphasizing before continuing. First, simple \textit{conservative} surrogates are still possible. For example, for $D\ge 0$, the following holds:
\begin{equation}
\psi(\bm{\chi},t)
=
A+B\cos t+C\sin t+D\cos^2t
\le
A+\sqrt{B^2+C^2}+D.
\label{eq:simple_surrogate_bound}
\end{equation}
Second, and critically, we emphasize that a pragmatic implementation of the non-Gaussian confidence boundary is to simply discretely sample along the boundary of the banana contour and jointly enforce inequality constraints on each point. The downsides are the potential for the sampling to be too sparse, and in the wasted computational effort in evaluating inequalities on many points, when a worst-violating or closest-violating point always exists. However, this approach works for preliminary implementations.

Solving for the worst-violating angle is numerically inexpensive compared to evaluating the constraint function or its gradients. 
We thus aim to first solve for this worst-case angle $t^{\star}$ maximizing Eq.~\eqref{eq:psi_ABCD}. A change of variables is convenient for this: 
\begin{equation}
\zeta \equiv \cos t, \qquad \zeta \in [-1,1].
\label{eq:zeta_def}
\end{equation}
\begin{equation}
\sin t = \pm \sqrt{1-\zeta^2},
\label{eq:sin_from_zeta}
\end{equation}
\begin{equation}
\psi_\pm({\bm{\chi}},\zeta)
=
A({\bm{\chi}})+B({\bm{\chi}})\zeta
\pm C({\bm{\chi}})\sqrt{1-\zeta^2}
+D({\bm{\chi}})\zeta^2.
\label{eq:psi_pm_zeta}
\end{equation}
The branch ambiguity is in the sign of the square-root term, and the maximizing sign is the one which matches the sign of $C({\bm{\chi}})$. Hence,
\begin{equation}
\max_t \psi({\bm{\chi}},t)
=
\max_{\zeta\in[-1,1]}
\left(
A({\bm{\chi}})+B({\bm{\chi}})\zeta+|C({\bm{\chi}})|\sqrt{1-\zeta^2}+D({\bm{\chi}})\zeta^2
\right).
\label{eq:zeta_support_reduction}
\end{equation}
Thus the continuous half-plane support computation reduces to a one-dimensional maximization over $\zeta\in[-1,1]$ for some maximizing $\zeta_\star$, 
which is trivially numerically solvable.

By definition, the following hold
\begin{subequations}
\begin{align}
\cos t^{\star}= & \ \zeta_\star \\
\sin t^{\star}
= & \ 
\operatorname{sgn}(C({\bm{\chi}}))\sqrt{1-\zeta_\star^2},
\qquad
C({\bm{\chi}})\neq 0.
\end{align}
\label{eq:sin_tstar}
\end{subequations}
Therefore
\begin{equation}
t^{\star}
=
\operatorname{atan2}
\!\Big(
\operatorname{sgn}(C({\bm{\chi}}))\sqrt{1-\zeta_\star^2},\ \zeta_\star
\Big),
\label{eq:tstar_recovery}
\end{equation}
which is unique whenever $C({\bm{\chi}})\neq 0$. When $C({\bm{\chi}})= 0$, the branching of Eq.~\eqref{eq:psi_pm_zeta} disappears and the solution for $t^{\star}$ is instead to maximize a quadratic in $\zeta$ then solve for $t^{\star}$. Once $t^{\star}$ is found, no other points on the contour are relevant to enforcing the constraint.

\paragraph*{Solution by Sequential Corrections:}
We seek values of the general problem free variables ${\bm{\chi}}$ minimizing some $J(\bm{\chi})$ while also satisfying the inequality constraint $g(\bm{\chi})\leq0$, where  $g(\bm{\chi}) = \underset{t}{\text{max}} \ \psi(\bm{\chi},t)=\psi(\bm{\chi},t^{\star})$. As this is a nonlinear problem, these are to be obtained by a local sequential correction procedure. When the constraint is violated, once the worst-violating contour point $t^{\star}$ is identified, a local correction can be defined. Consider for the sake of argument a simple first-order (gradient-based) local correction:
\begin{equation}
\Delta {\bm{\chi}}_{i} = -K_{1}\frac{g}{\|\nabla_{\chi}g\|^{2}}\,\nabla_{\bm{\chi}}g - K_{2}\left(I - \frac{\nabla_{\bm{\chi}}g\nabla_{\bm{\chi}}g^{\top}}{\|\nabla_{\chi}g\|^{2}}\right)\nabla_{\bm{\chi}} J,
\label{eq:delta_chi_update}
\end{equation}
for some $K_{1},K_{2}>0$ and iterate $i>1$, where $g$, $J$, $\nabla g$, and $\nabla J$ are evaluated on the prior iterate $\bm{\chi}_{i-1}$. 
The first term in Eq.~\eqref{eq:delta_chi_update} reduces the constraint violation, and the second term reduces the cost along an orthogonal direction. The associated gradient is $\nabla_{\bm{\chi}} g=\nabla_{\bm{\chi}}\psi(\bm{\chi},t^{\star})$:
\begin{equation}
\nabla_{\bm{\chi}} \psi({\bm{\chi}},t)
=
\nabla A({\bm{\chi}})
+\cos t\,\nabla B({\bm{\chi}})
+\sin t\,\nabla C({\bm{\chi}})
+\cos^2 t\,\nabla D({\bm{\chi}}).
\label{eq:grad_psi_ABCD}
\end{equation}
All gradients are defined in Appendix B. This is the gradient of the active half-plane residual evaluated at the worst contour point. As the worst violator $t^{\star}$ moves, holding $\bm{\chi}$ constant, the term $\nabla g$ updates as in Eq.~\eqref{eq:grad_psi_ABCD}, which affects corrective schemes such as Eq.~\eqref{eq:delta_chi_update}. Thus, we seek to understand how the nature of changes in $t^{\star}$ affects the stability of a sequential correction procedure.

\begin{figure}[htbp]
    \centering
    \subfigure[LinCov vs. Non-Gaussian Chance Constraint]{\includegraphics[width=0.52\textwidth]{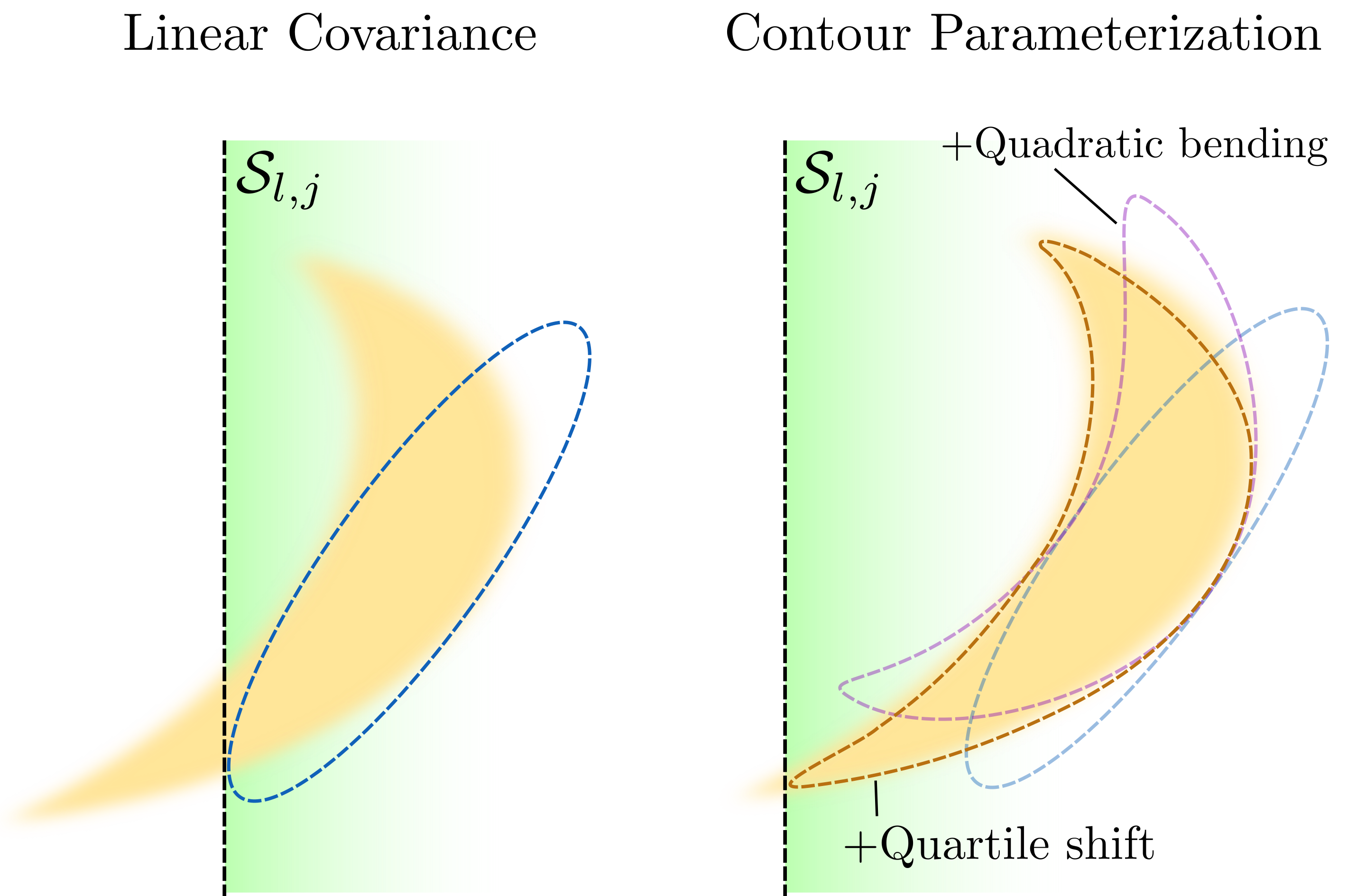}}
    \hfill
    \subfigure[Problematic Boundary Orientation]{\includegraphics[width=0.35\textwidth]{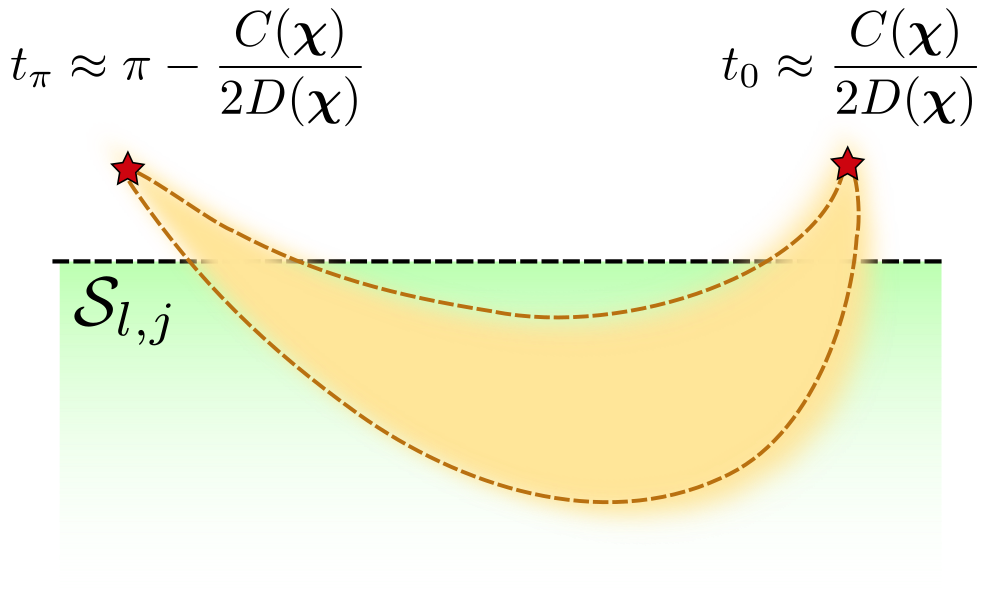}}
    \caption{Non-Gaussian Chance Constraint Geometry}
    \label{fig:badorient}
    \vspace{-1em}
\end{figure}

\paragraph*{Confidence contour non-convexity and constraint effects:}
For a coplanar half-plane constraint, the only possible non-smooth change in $t^{\star}$ in the present geometry is the competition between the two ``tips" of the banana contour. This occurs when the two ends of the confidence boundary are simultaneously equidistant or near equidistant to the half-plane constraint surface -- see Fig.~\ref{fig:badorient}(b). This is a concern because the corrective equation Eq.~\eqref{eq:delta_chi_update} is informed directly by the gradient of the constraint violation, Eq.~\eqref{eq:grad_psi_ABCD}, which itself is a trigonometric function of the maximum violating angle, holding ${\bm{\chi}}$ fixed. A non-smooth change in $t^{\star}$ from one iteration to the next thus invites the possibility of undesirable behaviors such as chatter or optimizer stall. It can be shown e.g. by inspection of Eqs.~\eqref{eq:banana_param_halfplane} that when the $D(\bm{\chi})$ term dominates, this occurs near the following angles:
\begin{equation}
t_{\text{crit}}\approx\{0, \ \pi\}, 
\label{eq:tip_angles}
\end{equation}

Let us elaborate further on this problematic orientation. The following holds:
\begin{equation}
B({\bm{\chi}})=k\,m_1({\bm{\chi}})\sqrt{\lambda_1({\bm{\chi}})},
\qquad
m_1({\bm{\chi}})=\bm{e}_1({\bm{\chi}})^\top \bm{n}.
\label{eq:B_geometry}
\end{equation}
A ``near-tie" in constraint violation between the two banana tips occurs only when $B({\bm{\chi}})\approx 0$ or $m_1({\bm{\chi}})=\bm{e}_1({\bm{\chi}})^\top \bm{n} \approx 0$, 
where $\bm{e}_{1}$ is the first basis vector of $R$. This is the degenerate alignment in which the half-plane normal is nearly orthogonal to the long principal axis. Only in the narrow regime of small $|B({\bm{\chi}})|$ can the active parameter jump non-smoothly between $t^{\star}\approx 0$ and $t^{\star}\approx \pi$. 

\subsection{Overcoming the Non-Convex Constraint}
\paragraph{I. Direct Gradient Smoothing:}
We must first identify exactly the $t_{\text{crit}}$ angles, and then propose a stable alternative to the discrete jump in gradient between these values. Starting from Eq.~\eqref{eq:psi_ABCD}, we 
first impose
$B(\bm{\chi})=0$. Then
\begin{equation}
\psi_0(\bm{\chi},t)=A(\bm{\chi})+C(\bm{\chi})\sin t+D(\bm{\chi})\cos^2 t.
\label{eq:psi0_def}
\end{equation}

Differentiating with respect to \(t\), the critical points can be identified as Eq.~\eqref{eq:critical_conditions}:
\begin{equation}
\psi_0'(\bm{\chi},t)
=
C(\bm{\chi})\cos t-D(\bm{\chi})\sin 2t
=
\cos t\Bigl(C(\bm{\chi})-2D(\bm{\chi})\sin t\Bigr).
\label{eq:dpsi0_dt}
\end{equation}
\begin{equation}
\cos t=0
\qquad\text{or}\qquad
\sin t=\frac{C(\bm{\chi})}{2D(\bm{\chi})}.
\label{eq:critical_conditions}
\end{equation}

The condition $C(\bm{\chi})/(2D(\bm{\chi})) < 1$ enables the second set. The two critical angles of interest, namely the ones lying on the two banana-tip branches near \(0\) and \(\pi\), are therefore
\begin{equation}
t_{0,\pi}(\bm{\chi})
=
\operatorname{atan2}\!\left(
\frac{C(\bm{\chi})}{2D(\bm{\chi})},
\pm\sqrt{1-\left(\frac{C(\bm{\chi})}{2D(\bm{\chi})}\right)^2}
\right).
\label{eq:t0pi_exact}
\end{equation}

Define the following useful identities:
\begin{equation}
s_*(\bm{\chi}):=\frac{C(\bm{\chi})}{2D(\bm{\chi})},
\qquad
c_*(\bm{\chi}):=\sqrt{1-s_*(\bm{\chi})^2}.
\label{eq:def_sstar_cstar}
\end{equation}
\begin{equation}
\sin t_0=\sin t_\pi=s_*,
\qquad
\cos t_0=+c_*,
\qquad
\cos t_\pi=-c_*,
\qquad
\cos^2 t_0=\cos^2 t_\pi=c_*^2.
\label{eq:tip_branch_values}
\end{equation}

Evaluating Eq.~\eqref{eq:psi_ABCD} at these two branch representatives and applying some trigonometry, 
\begin{equation}
\psi(\bm{\chi},t_0)=\Lambda(\bm{\chi})+B(\bm{\chi})c_*(\bm{\chi}),
\qquad
\psi(\bm{\chi},t_\pi)=\Lambda(\bm{\chi})-B(\bm{\chi})c_*(\bm{\chi}),
\label{eq:psi_tip_pm}
\end{equation}
where
\begin{equation}
\Lambda(\bm{\chi})
:=
A(\bm{\chi})+D(\bm{\chi})+\frac{C(\bm{\chi})^2}{4D(\bm{\chi})}.
\label{eq:def_Lambda}
\end{equation}

The ``log-sum-exp" function (see e.g. Ref.~\citenum{Boyd_Vandenberghe_2004}) gives a surrogate
\begin{equation}
g_{\mathrm{loc},\tau}(\bm{\chi})
:=
\tau\log\!\left(
e^{\psi(\bm{\chi},t_0)/\tau}+e^{\psi(\bm{\chi},t_\pi)/\tau}
\right)
\label{eq:def_g_loc_tau_clean}
\end{equation}
which smoothly recovers the true maximum of $\psi(\bm{\chi},t)$ at $t_{0}$ and $t_{\pi}$ as $\tau\rightarrow 0^{+}$, and thus acts as a local proxy for $\psi$ in the problematic orientation. Substituting Eq.~\eqref{eq:psi_tip_pm} yields
\begin{equation}
g_{\mathrm{loc},\tau}(\bm{\chi})
=
\Lambda(\bm{\chi})+\tau\log\!\left(
2\cosh\!\frac{B(\bm{\chi})c_*(\bm{\chi})}{\tau}
\right).
\label{eq:g_loc_final_clean}
\end{equation}

\noindent\textbf{\underline{Key Result 2}:} 
For a single half-plane chance constraint $g$, in lieu of a discontinuous change in $\nabla_{\bm{\chi}}g$ between $t_{0}$ and $t_{\pi}$, we seek a stable alternative. Differentiating Eq.~\eqref{eq:g_loc_final_clean} gives a smooth transition of the gradient in the vicinity of the problem orientation near $|B(\bm{\chi})|=0$.
\begin{equation}
\nabla_{\bm{\chi}} g_{\mathrm{loc},\tau}(\bm{\chi})
=
\nabla_{\bm{\chi}} \Lambda(\bm{\chi})
+
\tanh\!\left(
\frac{B(\bm{\chi})c_*(\bm{\chi})}{\tau}
\right)
\nabla_{\bm{\chi}}\!\bigl(B(\bm{\chi})c_*(\bm{\chi})\bigr).
\label{eq:grad_g_loc_final_clean}
\end{equation}
This proxy can be implemented whenever $|B(\bm{\chi})|$ is small, with judicious choice of smoothing parameter $\tau$. Away from the problematic orientation, the normal gradient can be used. All necessary gradients defined in Appendix B.

For small \(\left|C(\bm{\chi})/(2D(\bm{\chi}))\right|\), the two critical angles remain close to \(0\) and \(\pi\), with
\begin{equation}
t_0(\bm{\chi})\approx \frac{C(\bm{\chi})}{2D(\bm{\chi})},
\qquad
t_\pi(\bm{\chi})\approx \pi-\frac{C(\bm{\chi})}{2D(\bm{\chi})}.
\label{eq:small_ratio_angles}
\end{equation}

\paragraph{II. Conservative Surrogate:}
Previously we discussed solving for $t^{\star}$ maximizing $\psi(\bm{\chi},t)$, and enforcing $\psi(\bm{\chi},t^{\star})\leq0$ to satisfy the chance constraint, noting that this process does not produce smooth corrective updates $\Delta\bm{\chi}$ on problem free variables. We can obviate the need for computing $t^{\star}$ entirely via a more elegant but more complex approach, by defining a tunable smooth surrogate for the non-convex constraint. For a fixed value of $\bm{\chi}$, define the maximum angular value of $\psi$:
\begin{equation}
M := \max_{t\in[0,2\pi]} \psi(t).
\label{eq:def_M}
\end{equation}
The ``log-integral-exp'' surrogate, inspired by log-sum-exp, approaches $M$ as $\tau\rightarrow 0^+$:
\begin{equation}
g_{\text{b},\tau}[\psi]
:=
\tau \log\!\left(
\frac{1}{2\pi}\int_{0}^{2\pi} e^{\psi(t)/\tau}\,dt
\right),
\qquad \tau>0,
\label{eq:def_g_tau}
\end{equation}
where the ``b'' subscript denotes ``below''. One can show that Eq.~\eqref{eq:def_g_tau} always satisfies $g_{\text{b},\tau}[\psi] \le M$. Thus, $g_{\text{b},\tau}[\psi]$ alone is not a conservative replacement for Eq.~\eqref{eq:def_M}. 

We need a conservative smooth \textit{upper bounding} function. Define the maximum perimetric derivative $L$, noting $\psi$ is smooth and analytic, and $( \ )^{'}=\frac{\text{d}}{\text{d}t}( \ )$:
\begin{equation}
L := \max_t |\psi'(t)|.
\label{eq:def_L}
\end{equation}
As shown in Appendix C, the following conservative upper bound holds:
\begin{equation}
\max_t\psi(t)
\le
g_{\text{b},\tau}[\psi] + \mathcal{C}(\tau,L),
\label{eq:final_upper_bound}
\end{equation}
where
\begin{equation}
\mathcal{C}(\tau,L)
:=
-\tau\log\!\left(
\frac{\tau}{2\pi L}\left(1-e^{-2\pi L/\tau}\right)
\right).
\label{eq:def_C}
\end{equation}
This term satisfies $\mathcal{C}(\tau,L)\ge 0$ and furthermore, $\mathcal{C}(\tau,L)\to 0$ as $\tau\to 0^+$.

\begin{figure}[h!]
\centering
\includegraphics[scale=0.99]{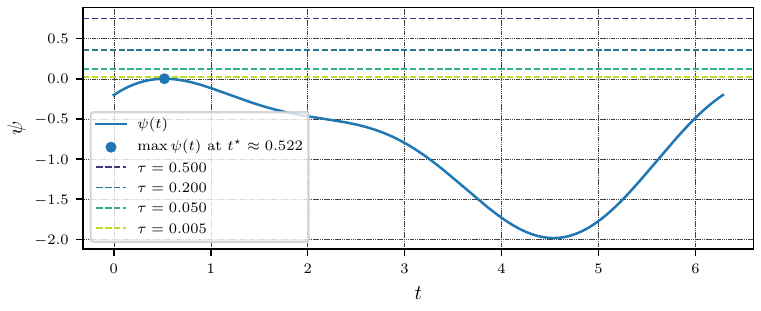}
\caption{Example Demonstration of Conservative Surrogate for Various Values of $\tau$}
\label{fig:softmax}
\vspace{-1em}
\end{figure}

The original constraint $\psi(t^{\star})\leq0$ is conservatively enforced by the following condition, which side-steps the need to solve for the worst violating angle $t^{\star}$:
\begin{equation}
g_{\text{b},\tau}[\psi] + \mathcal{C}(\tau,L)\le 0.
\label{eq:conservative_smooth_constraint}
\end{equation}
Writing the conservative upper bound as $g_{\text{a},\tau}[\psi] := g_{\text{b},\tau}[\psi] + \mathcal{C}(\tau,L)$, where subscript ``a'' denotes ``above'', we observe that enforcing $g_{\text{a},\tau}[\psi]\leq0$ conservatively satisfies Eq.~\eqref{eq:support_constraint_halfplane}. Explicitly, Eq.~\eqref{eq:support_constraint_halfplane} is replaced by a new function of the decision variables $\bm{\chi}$. 
This is just a new $g$ function in Eq.~\eqref{eq:delta_chi_update}, for which we need to compute the necessary $\nabla_{\bm{\chi}}g$ terms. An example plot of $g_{\text{a},\tau}[\psi]$ is given in Fig.~\ref{fig:softmax}, illustrating the effect of tuning parameter $\tau$.

\noindent\textbf{\underline{Key Result 3}:} The aforementioned tunable conservative chance constraint surrogate $g_{\text{a}}$ is smooth in decision variables $\bm{\chi}$ for all practical purposes, as is its gradient:
\begin{equation}
\nabla_{\bm{\chi}}g_{\text{a},\tau}[\psi] \approx \nabla_{\bm{\chi}}g_{\text{b},\tau}[\psi],
\qquad
\tau \to 0^+.
\label{eq:grad_integral_surrogate}
\end{equation}
As discussed in Appendix C, the parameter $L$ may be safely treated as a \textit{fixed} conservative parameter, or updated occasionally. All necessary gradients are defined in Appendices B and C. When properly tuned, this approach offers strong guarantees of performance, but the equations for $g_{\text{a},\tau}$ and its gradient contain integrals which may become numerically burdensome to compute as $\tau\rightarrow 0^{+}$. 

\section{Numerical Results}
\subsection{Remark on Higher-Order Moment Computations}
The non-Gaussian confidence contour parameterization given by Eq.~\eqref{eq:banana_param_halfplane} requires that we estimate the skew and kurtosis of the distribution. This can be accomplished in a number of ways of varying efficiency. Importantly, because the resulting confidence contour is used in an optimizer, we need a deterministic method for estimating the statistical moments. We use the Conjugate Unscented Transform (CUT)\cite{CUT_ACC}, which generally has significant speed advantages over random sampling-based methods. The equations for implementation are provided in the Appendix for completeness. See Refs.~\citenum{AAS26b}, \citenum{BurnettBoone_ISSFD26} for discussions of different techniques for estimating higher-order statistical moments. In this paper, all numerical results are obtained from Python scripts using open-source optimization tools like \texttt{SLSQP} from \texttt{scipy.optimize}, on a 2024 MacBook Pro with Apple M4 Max chip.
\subsection{I. Numerical Comparison of Different Methodologies: An Asteroid Orbiter Example}
\begin{figure}[h!]
\vspace{-1em}
\begin{center}
\includegraphics[scale=0.7]{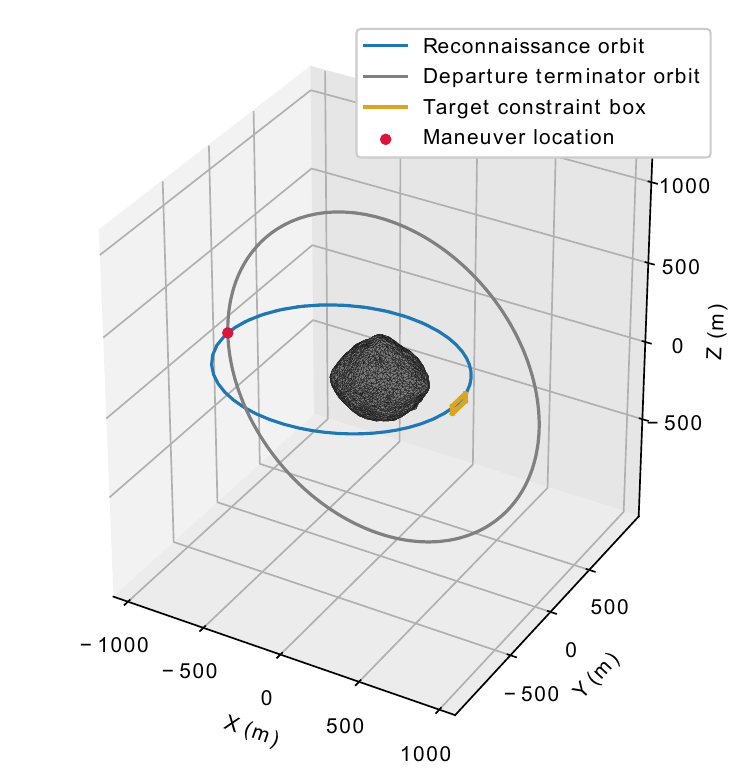}
\caption{Asteroid maneuver targeting scenario}
\label{f:asteroid_scenario_shorter}
\end{center}
\vspace{-1em}
\end{figure}

As a first test of the stochastic control techniques discussed previously, we repeat the example of Ref.~\citenum{boone_cdc} which was originally solved with a Gaussian mixture model. The spacecraft is initialized about an asteroid with gravitational parameter $\mu=5.2~\mathrm{m}^3/\mathrm{s}^2$ with the state
\[
\bm{x}_0 =
\begin{bmatrix}
-1000 & 0 & 0 & 0 & 0 & -7.211\times10^{-2}
\end{bmatrix}^{T}
\]
with position and velocity components in m, m/s respectively. At the initial time $t_0$, an impulsive control $\bm{u}_0$ is selected to transfer the spacecraft toward a reconnaissance trajectory that descends closer to the asteroid surface, as illustrated in Fig.~\ref{f:asteroid_scenario_shorter}. The maneuver is designed subject to a terminal safety requirement at time $t_1$: despite uncertainty in the initial state, the propagated spacecraft distribution should not approach the asteroid more closely than the prescribed keep-out limit. 

The initial uncertainty is taken to be Gaussian with diagonal covariance $P_0=\mathrm{diag}(\bm{\sigma}_0^2)$, where
\begin{equation}
\bm{\sigma}_0 =
\begin{bmatrix}
1.0 & 1.0 & 1.0 & \epsilon & \epsilon & \epsilon
\end{bmatrix}^{T}
\quad
(\mathrm{m},\mathrm{m/s}),
\end{equation}
and $\epsilon\ll1$. This corresponds to a position-dominated dispersion, which is appropriate for the present problem because the escape speed is only on the order of $0.1~\mathrm{m/s}$. The terminal time $t_1$ is chosen to occur after 1.5 revolutions of the reconnaissance orbit following the maneuver, approximately $23.6$ hours later. By this time, the initially Gaussian uncertainty has evolved into a visibly non-Gaussian distribution in Cartesian components. The scenario is depicted in Figure~\ref{f:asteroid_scenario_shorter}.

\begin{figure}[h!]
\centering
\subfigure[LinCov constraint\label{f:linear_plot}]{
\includegraphics[width=0.49\textwidth]{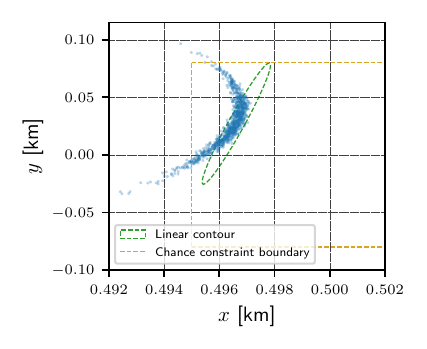}}
\hfill
\subfigure[Banana contour constraint\label{f:banana_plot}]{
\includegraphics[width=0.49\textwidth]{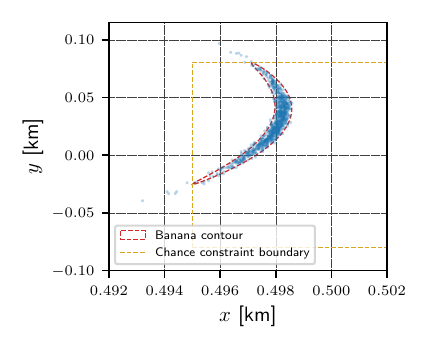}}
\caption{Monte Carlo Outcomes of LinCov vs. Banana Stochastic Control}
\label{f:contour_constraint_comparison}
\vspace{-1em}
\end{figure}

The terminal requirement is imposed as a rectangular keep-in box in position space, represented by six scalar chance constraints below. Each face is assigned the risk allocation, $\Delta_1=0.01$, so each corresponding constraint must be satisfied by at least $99\%$ of the distribution.
\begin{subequations}
\begin{align}
& \textrm{p} [x_1 \geq 495 \textrm{ m}] \geq 1 - \Delta_1 \\
& \textrm{p} [x_1 \leq 505 \textrm{ m}] \geq 1 - \Delta_1 \\
& \textrm{p} [y_1 \geq -80 \textrm{ m}] \geq 1 - \Delta_1 \\
& \textrm{p} [y_1 \leq 80 \textrm{ m}] \geq 1 - \Delta_1 \\
& \textrm{p} [z_1 \geq -25 \textrm{ m}] \geq 1 - \Delta_1 \\
& \textrm{p} [z_1 \leq 25 \textrm{ m}] \geq 1 - \Delta_1
\label{eq:cc_box}
\end{align}
\end{subequations}
Figure~\ref{f:contour_constraint_comparison} shows the result of banana and LinCov policies in a 5000 run Monte Carlo simulation. The LinCov solution satisfies the keep-out constraint in $88.10\%$ of samples, while the banana solutions improve the satisfaction rate to about $98.35\%$. Note each \textit{individual} constraint is successfully satisfied to 99.0\% probability. 
Both methods require similar control of  $\|\bm{u}\|\approx9.3\times10^{-2}~\mathrm{m/s}$.

Table~\ref{tab:asteroid_runtime_comparison} gives the optimizer runtime for the asteroid targeting problem for all major implementation strategies discussed in this paper. Here $g$ denotes a constraint-function callback and $Dg$ denotes a constraint-derivative callback. Entries marked FD use finite-difference constraint derivatives, so derivative information is obtained by repeated calls to $g$. The analytic variants instead provide separate $Dg$ callbacks. The timing column $t_{\rm UQ}$ denotes the portion of the optimizer runtime spent propagating the covariance or higher-order uncertainty description inside the constraint evaluations.
The table shows that the different variants of the non-Gaussian stochastic maneuver design approach have similar runtimes despite different callback counts. The sampled contour method requires the most constraint function calls. The active-angle and integral variants reduce the number of constraint function calls by supplying analytic derivatives, but the derivative-enabled uncertainty construction is more expensive per evaluation. These effects largely offset one another. Thus, for this example, the active-angle and integral surrogate formulations improve the mathematical structure of the constraint enforcement, but they do not substantially reduce optimizer runtime relative to a contour sampling approach. 

\begin{table}[h!]
\vspace{1em}
\centering
\caption{Runtime breakdown for asteroid targeting optimizations}
\label{tab:asteroid_runtime_comparison}
\renewcommand{\arraystretch}{1.12}
\begin{tabular}{@{}lcccc@{}}
\toprule
Method & Derivatives & Calls & $t_{\rm opt}$ [s] & $t_{\rm UQ}$ [s] \\
\midrule
LinCov & FD & $23g$ & $0.604$ & $0.598$ \\
Banana sampled & FD & $18g$ & $9.567$ & $9.097$ \\
Banana active & analytic & $6g+3Dg$ & $9.018$ & $8.377$ \\
Banana integral & analytic & $6g+3Dg$ & $8.950$ & $8.335$ \\
\bottomrule
\end{tabular}
\end{table}

\subsection{II. A Full Stochastic Guidance Problem: Lunar Free-Return Midcourse Corrections}
We now introduce an applied example motivated by the recent Artemis II mission, which marked humanity's first crewed visit to the Moon in over 50 years. The example is based on a simple planar circular restricted three-body problem (CR3BP, see Ref.~\citenum{Koon:2006rf}) analog of the Artemis II free-return trajectory. 
Our purpose here is not to re-optimize the Artemis II ConOps, but rather to give a simple preliminary demonstration of how the proposed non-Gaussian chance-constraints might be applied in long-horizon spacecraft targeting. For greater realism, Reference~\citenum{WoffindenAAS2023} provides an actual NASA study of Artemis II stochastic guidance using a LinCov framework. 

\begin{figure}[h!] 
\centering
\subfigure[Inertial Frame\label{fig:artemis_inertial}]{
\includegraphics[width=0.45\textwidth]{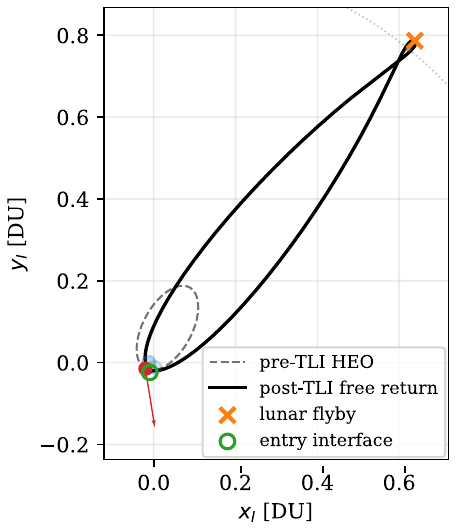}}
\hspace{1em}
\subfigure[Rotating Frame\label{fig:artemis_rotating}]{
\includegraphics[width=0.5\textwidth]{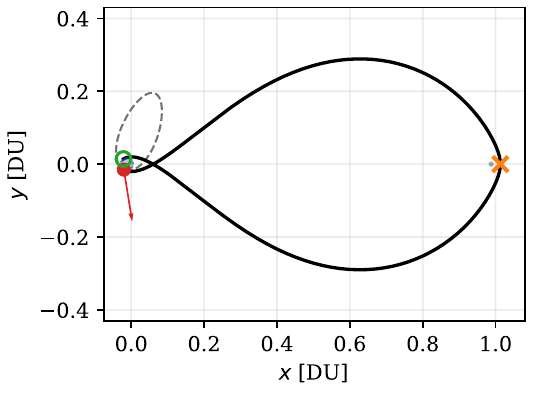}}
\caption{Artemis II-like Planar Free Return}
\label{fig:CC1}
\vspace{-1em}
\end{figure}

Figure~\ref{fig:CC1} shows the corresponding nominal planar free-return trajectory. 
Consider hypothetical midcourse correction maneuvers performed on the post-lunar-flyby return leg. 
The navigation state error and maneuver execution error effects are treated statistically. Let
\begin{equation}
    \bm{x}_m^- =
    \hat{\bm{x}}_m^- + \delta\bm{x}_m^-,
    \qquad
    \delta\bm{x}_m^- \sim \mathcal{N}(\bm{0},P_m^-),
    \label{eq:post_nav_midcourse_state}
\end{equation}
where \(\hat{\bm{x}}_m^-\) is the post-navigation estimate immediately before the correction maneuver. Let \(\Delta\bar{\bm{v}}_m\) denote the commanded correction maneuver, and model the executed maneuver as
\begin{equation}
    \Delta\bm{v}_m =
    \Delta\bar{\bm{v}}_m + \bm{e}_{\Delta v},
    \qquad
    \bm{e}_{\Delta v}\sim \mathcal{N}(\bm{0},Q_{\Delta v}).
    \label{eq:midcourse_execution_error}
\end{equation}
For an impulsive maneuver,
\begin{equation}
    \bm{x}_m^+
    =
    \bm{x}_m^-
    +
    B_m\Delta\bar{\bm{v}}_m
    +
    B_m\bm{e}_{\Delta v},
    \label{eq:post_cleanup_state}
\end{equation}
where \(B_m = [0_{3\times 3}, I_{3\times 3}]^{\top}\). Therefore, the
post-maneuver nominal state and covariance are
\begin{equation}
    \hat{\bm{x}}_m^+
    =
    \hat{\bm{x}}_m^-
    +
    B_m\Delta\bar{\bm{v}}_m,
    \label{eq:post_cleanup_nominal}
\end{equation}
\begin{equation}
    P_m^+
    =
    P_m^-
    +
    B_m Q_{\Delta v} B_m^\top.
    \label{eq:post_cleanup_covariance}
\end{equation}

The classical JPL Gates error model\cite{Gates1963} provides a simplified description of maneuver execution error via distinct modalities of shutoff, resolution, pointing, and autopilot errors. The execution covariance is decomposed into components parallel and perpendicular to the commanded maneuver, based on the various error pathways and statistics. Defining $V = \|\Delta\bar{\bm{v}}_m\|$ and $\hat{\bm{e}} = \frac{\Delta\bar{\bm{v}}_m}{\|\Delta\bar{\bm{v}}_m\|}$, we write
\begin{equation}
    Q_{\Delta v}
    =
    \left(\sigma_r^2 + V^2\sigma_s^2\right)\hat{\bm{e}}\hat{\bm{e}}^\top
    +
    \left(\sigma_a^2 + V^2\sigma_p^2\right)
    \left(I-\hat{\bm{e}}\hat{\bm{e}}^\top\right).
    \label{eq:gates_projector_form}
\end{equation}
Here \(\sigma_r\) is a fixed along-burn resolution error, \(\sigma_s\) is a
fractional shutoff or scale error, \(\sigma_p\) is a pointing error in radians,
and \(\sigma_a\) is a fixed transverse autopilot error. See Reference~\citenum{Gates1963} for more information. 
Thus the cleanup command affects the stochastic problem both by shifting the
nominal post-maneuver state and by changing the state covariance
through the commanded nominal maneuver. The downstream uncertainty propagation is
then posed as
\begin{equation}
    \bm{x}_f
    =
    \bm{\varphi}_{t_f,t_m}
    \left(
    \hat{\bm{x}}_m^+ + \delta\bm{x}_m^+
    \right),
    \qquad
    \delta\bm{x}_m^+\sim \mathcal{N}(\bm{0},P_m^+),
    \label{eq:post_cleanup_downstream_uq}
\end{equation}
where \(\bm{\varphi}_{t_f,t_m}\) denotes the flow of the CR3BP dynamics from the
cleanup epoch to the atmospheric entry interface. As these dynamics are nonlinear, the final state distribution at the atmospheric entry point will no longer be Gaussian. Critically, we emphasize the following point: 
For passive safety after the post-lunar-flyby correction maneuver, this maneuver must account for the expected final non-Gaussian uncertainty, even if navigation updates continue until
atmospheric entry or until the next burn. Later navigation updates may reduce epistemic uncertainty in the estimated state, but they do not reduce the physical dispersion of outcomes induced by the preceding correction maneuver.

The entry constraints are in local entry coordinates. Let \(\bm{r}\) denote the
Earth-relative position at entry and let \(\bm{v}\) denote the relative inertial velocity. Defining \(\hat{\bm{e}}_r = \bm{r}/\|\bm{r}\|\), decompose the velocity:
\begin{equation}
    v_R = \hat{\bm{e}}_r^\top \bm{v},
    \qquad
    v_T = \hat{\bm{e}}_t^\top \bm{v},
    \label{eq:local_entry_velocity_components}
\end{equation}
where \(\hat{\bm{e}}_t\) is the local tangential direction. The entry
flight-path angle satisfies \(\tan\gamma = v_R/v_T\), assuming \(v_T>0\).
Rather than constraining \(\gamma\) directly, the shallow- and steep-side entry
constraints can be written as half-plane constraints in the local velocity plane. Table~\ref{tab:edl-corridor-halfplanes} gives a compact form of the entry
corridor constraints. The two position constraints enforce an admissible
entry-interface radial window, while the two velocity constraints enforce
shallow and steep flight-path-angle limits. 

\begin{table}[h!]
\centering
\caption{Lunar Return Earth Arrival/Re-entry Constraints}
\label{tab:edl-corridor-halfplanes}
\renewcommand{\arraystretch}{1.25}
\begin{tabular}{lll}
\toprule
Constraint type & Generic form & Local EDL corridor version \\
\midrule
Position half-plane ($\times2$)
&
$\bm{n}_r^\top \bm{r} - b_0 \le 0$
&
$\hat{\bm{e}}_r^\top \bm{r} - r_{\max} \le 0$
\\
&
&
$-\hat{\bm{e}}_r^\top \bm{r} + r_{\min} \le 0$
\\
Velocity half-plane ($\times2$)
&
$\bm{n}_v^\top \bm{v} - b_0 \le 0$
&
$v_R - \tan(\gamma_{\mathrm{shallow}})v_T \le 0$
\\
&
&
$-v_R + \tan(\gamma_{\mathrm{steep}})v_T \le 0$
\\
\bottomrule
\end{tabular}
\vspace{-0.5em}
\end{table}

We consider a hypothetical sequence of two correction maneuvers during mission operations to re-target the entry corridor, computing a policy with high statistical confidence of success. The maneuver error parameters are given in Table~\ref{tab:Gates}, with a fairly pessimistic proportional error of 3\%. In this example, the nominal re-entry condition has entry flight path angle (EFPA) of $\gamma=-6^{\circ}$ and altitude of $h=120$ km, but state errors after the lunar flyby are sufficient that trajectory correction is necessary. The first scheduled maneuver occurs shortly after lunar flyby, anticipating significant error and dispersion at the scheduled re-entry time due to navigation and maneuver errors. These error parameters are given in Table~\ref{tab:ErrorArtemis}. Out-of-plane errors are assumed subdominant for this planar example. In this context, due to significant expected dispersions, the final distribution is non-Gaussian, and exactly re-targeting the nominal re-entry condition is not \textit{passively safe}: if a follow-up maneuver cannot be completed in time, the spacecraft is at risk of burning up in the Earth's atmosphere from a too-steep entry condition. We seek instead a passively safe policy for the first maneuver, which corrects the expected dispersion to be closer to the nominal entry condition, while 
enforcing four three-sigma chance constraints
\begin{subequations}
\label{CCArtemisII}
\begin{align}
    \textrm{p}[h > 150~\mathrm{km}] &\ge \Phi(3) = 0.99865, \\
    \textrm{p}[h < 400~\mathrm{km}] &\ge \Phi(3) = 0.99865, \\
    \textrm{p}[\gamma > -6^\circ] &\ge \Phi(3) = 0.99865, \\
    \textrm{p}[\gamma < 8^\circ] &\ge \Phi(3) = 0.99865.
\end{align}
\end{subequations}
The lower constraints on $h$ and $\gamma$ are operationally critical. The upper constraints are enforced simply to prevent solutions that are ``safe" but arbitrarily far from the nominal entry condition.

\begin{table}[h!] \centering \caption{Maneuver Execution Error Parameters} \label{tab:Gates} \renewcommand{\arraystretch}{1.0} \begin{tabular}{lll} \toprule Parameter & Interpretation & Value \\ \midrule $\sigma_s$ & Proportional magnitude/shutoff error & $3.0\times 10^{-2}$ \\ $\sigma_r$ & Fixed error along the commanded burn direction & $10.0~\mathrm{mm/s}$ \\ $\sigma_p$ & Pointing error standard deviation & $3.0\times 10^{-4}~\mathrm{rad}$ \\ $\sigma_a$ & Fixed transverse execution error & $0.9~\mathrm{mm/s}$ \\ 
\bottomrule \end{tabular}
\vspace{-0.5em}
\end{table}

\begin{table}[h!]\centering \caption{Artemis Lunar Return Error: Correction Burn 1}\label{tab:ErrorArtemis} \renewcommand{\arraystretch}{1.25} \begin{tabular}{lll}\toprule Quantity & Key & Value \\\midrule 
Midcourse maneuver 1 time & $t_{m_{1}}$ & 0.25 days post-flyby \\
Time to entry & $t_{f} - t_{m_{1}}$ & 3.6163 days \\
Mean cleanup state error & $\delta\bm{x}_c = (\delta R,\delta T,\delta V_R,\delta V_T)$ & $(-0.5,50.0,-0.1,5.0)$ \\ & Units & $[\mathrm{km},\mathrm{km},\mathrm{m/s},\mathrm{m/s}]$ \\ Nav. position uncertainty & $(\sigma_R,\sigma_T,\sigma_N)$ & $(500.0,5000.0,5.0)~\mathrm{m}$ \\ Nav. velocity uncertainty & $(\sigma_{V_R},\sigma_{V_T},\sigma_{V_N})$ & $(50.0,500.0,0.5)~\mathrm{mm/s}$ \\ \bottomrule \end{tabular} 
\vspace{-0.5em}
\end{table}

\begin{figure}[h!]
\centering
\subfigure[Full trajectory with Correction Burn 1\label{fig:cleanup_traj_lincov}]{
\includegraphics[width=0.48\textwidth]{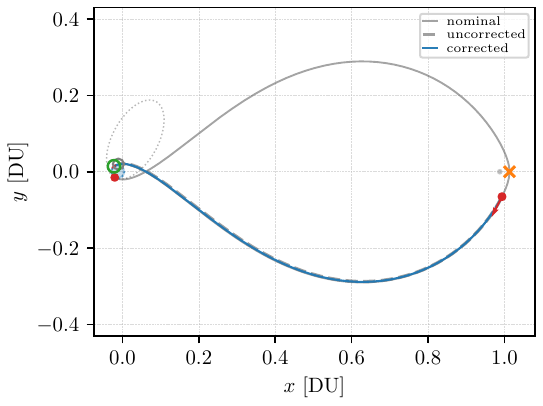}}
\hfill
\subfigure[Zoom with final 3$\sigma$ position bounds\label{fig:cleanup_traj_banana}]{
\includegraphics[width=0.48\textwidth]{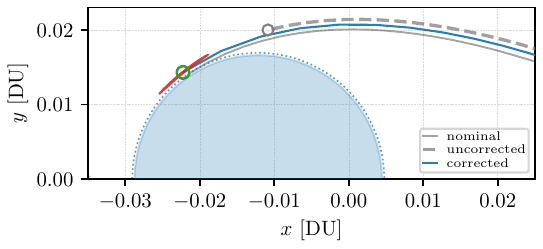}}
\caption{Post-flyby stochastic maneuver planning with banana policy: Correction Burn 1}
\label{fig:cleanup_trajectories}
\vspace{-1em}
\end{figure}

\begin{figure}[h!] \centering \subfigure[LinCov policy\label{fig:cleanup_pos_lincovA}]{ \includegraphics[width=0.48\textwidth]{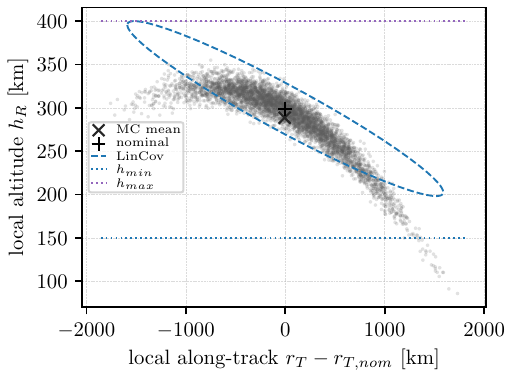}} \hfill \subfigure[Banana policy\label{fig:cleanup_pos_bananaA}]{ \includegraphics[width=0.48\textwidth]{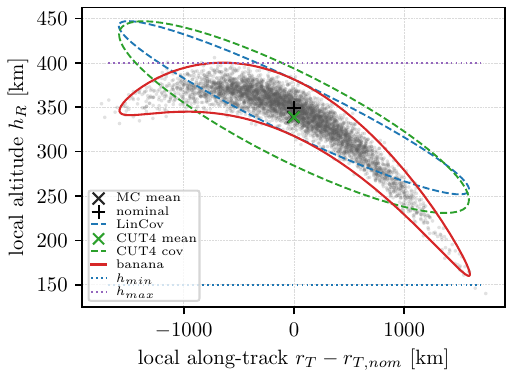}} \caption{Arrival position constraint satisfaction with Monte Carlo} \label{fig:cleanup_position_constraints} 
\end{figure}

\begin{figure}[h!] \centering \subfigure[LinCov policy\label{fig:cleanup_vel_lincov}]{ \includegraphics[width=0.48\textwidth]{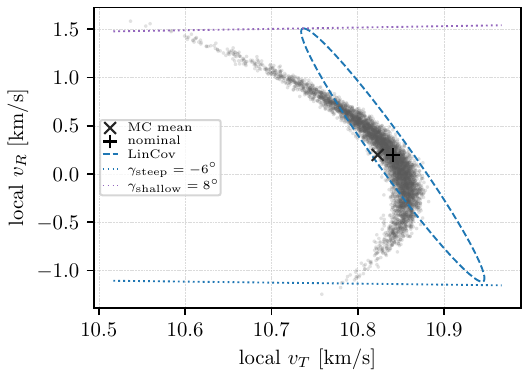}} \hfill \subfigure[Banana policy\label{fig:cleanup_vel_banana}]{ \includegraphics[width=0.48\textwidth]{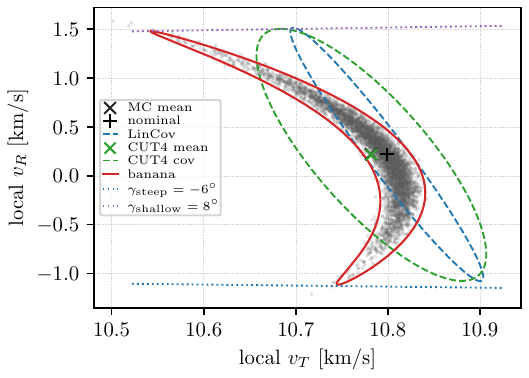}} \caption{Arrival velocity constraint satisfaction with Monte Carlo} \label{fig:cleanup_velocity_constraints}
\vspace{-1em}
\end{figure}

A naive maneuver, directly targeting a desired $\gamma=-6^{\circ}$ and $h=120$ km, requires $6.485~\mathrm{m/s}$ and places the nominal state directly on the lower-altitude/steep-entry boundary, which is not passively safe. In a 5000 sample Monte Carlo check, this violates the constraints in $62.86\%$ of samples, dominated by unacceptable low altitude and steep flight-path angles at the desired arrival time. The LinCov policy requires $3.352~\mathrm{m/s}$ and lowers the Monte Carlo violation percentage to 
0.56\%, but residual violations remain because the true final dispersion is not well-accounted for by the LinCov ellipse. See Figs.~\ref{fig:cleanup_position_constraints} and~\ref{fig:cleanup_velocity_constraints} for both LinCov and banana policy outcomes. The dispersion plots for the latter also show naive linear and nonlinear covariance ellipse predictions along with the banana contour for the same computed statistics. The banana policy, which is warm-started with the LinCov solution, reduces the maneuver from the LinCov solution to $2.867~\mathrm{m/s}$ while lowering the violation percentage to 0.2\%. 
The resulting trajectory is given in Fig.~\ref{fig:cleanup_trajectories} with the final 3$\sigma$ position bounds at the final time. Relative to LinCov, the banana policy saves $0.484~\mathrm{m/s}$, or $14.4\%$, and reduces the observed violation fraction by approximately $64\%$. We note that the banana-shaped confidence boundary fits the final Monte Carlo samples extremely well, even though it only makes use of LinCov-computed covariance and skew/kurtosis components from CUT4. The method does not use Monte Carlo sampling at all. The solve times in a Python script are 0.2s for naive retargeting, 0.25s for LinCov, and 4.29s using the banana solver. While more numerically expensive than LinCov, the approach has $\sim$$84\%$ lower runtime than the Monte Carlo study, as 5000 samples took 27.6 seconds in this example.

We now move on to the second corrective maneuver in our two-burn study. Key parameters given in Table~\ref{tab:ErrorArtemis2}, with the mean state error reported with respect to the nominal outcome of maneuver 1. Out-of-plane errors are kept small for this planar example. The proportional maneuver execution error $\sigma_{s}$ is relaxed from its prior pessimistic stress-test value to $5\times 10^{-3}$. All other parameters in Table~\ref{tab:Gates} are unchanged. The navigation estimate at maneuver 2 time is chosen as a representative outcome from maneuver 1, close to the boundary of $1\sigma$ dispersed outcomes from maneuver 1. 

\begin{table}[h!]\centering
\caption{Artemis Lunar Return Error: Correction Burn 2}\label{tab:ErrorArtemis2}
\renewcommand{\arraystretch}{1.0}
\begin{tabular}{lll}\toprule
Quantity & Key & Value \\\midrule 
Midcourse maneuver 2 time & $t_{m_{2}}$ & 5 hr before Earth entry interface \\
Time to entry & $t_{f} - t_{m_{2}}$ & $0.2083$ days \\
Mean cleanup state error & $\delta\bm{x}_c = (\delta R,\delta T,\delta V_R,\delta V_T)$ & $(-3.50,-40.78,-0.0094,0.500)$ \\
& Units & $[\mathrm{km},\mathrm{km},\mathrm{m/s},\mathrm{m/s}]$ \\
Nav. position uncertainty & $(\sigma_R,\sigma_T,\sigma_N)$ & $(333.3,1333.3,33.3)~\mathrm{m}$ \\
Nav. velocity uncertainty & $(\sigma_{V_R},\sigma_{V_T},\sigma_{V_N})$ & $(16.7,83.3,1.67)~\mathrm{mm/s}$ \\
\bottomrule
\end{tabular} 
\vspace{-0.5em}
\end{table}
\begin{figure}[h!]
\begin{center}
\includegraphics[width=0.61\textwidth]{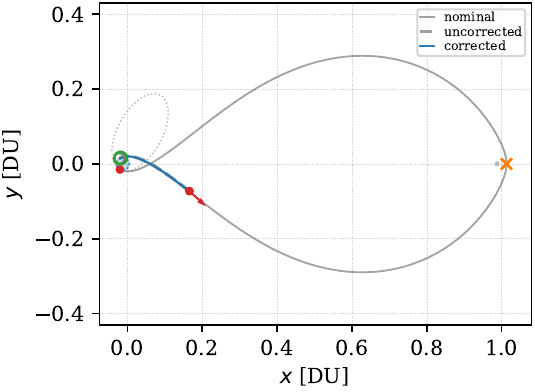}
\caption{Pre-entry stochastic maneuver planning with banana policy: Correction Burn 2}
\label{fig:cleanup_trajectories2}
\end{center}
\vspace{-1em}
\end{figure}
\begin{figure}[h!] \centering \subfigure[Position\label{fig:cleanup_pos_lincovB}]{ \includegraphics[width=0.48\textwidth]{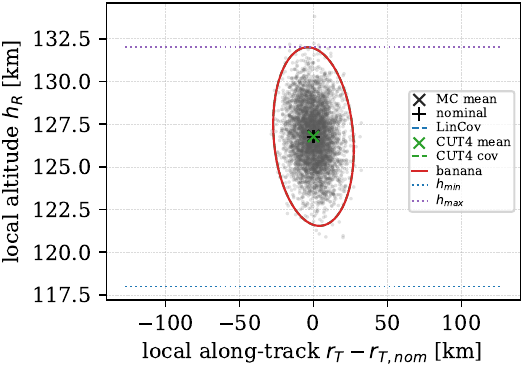}} \hfill \subfigure[Velocity\label{fig:cleanup_pos_bananaB}]{ \includegraphics[width=0.48\textwidth]{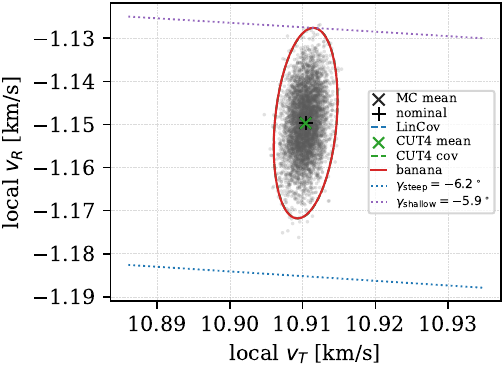}} \caption{Arrival constraint satisfaction with Monte Carlo (banana policy)} \label{fig:cleanup_constraints2} 
\vspace{-1em}
\end{figure}

For the second correction maneuver, the final entry corridor is enforced, again in one-sided inequalities at $3\sigma$ confidence, as the union of altitude range $118 < h < 132$ km and EFPA range $-6.2^{\circ} < \gamma < -5.9^{\circ}$. Entry longitude is allowed to vary in this example.
The resulting trajectory is given in Fig.~\ref{fig:cleanup_trajectories2}. As this maneuver executes with significantly lower navigation error and less propagation time, the final dispersion is much closer to Gaussian, so the LinCov and banana policies are essentially the same, with delta-V of 61.37 m/s, and Monte Carlo violation percentage of 0.14\%. The final position and velocity confidence bounds and dispersions obtained with the banana policy are provided in Figure~\ref{fig:cleanup_constraints2}. The bounds are accurate and respect the desired corridor with some margin. This second maneuver illustrates that the banana policy converges to a LinCov policy for Gaussian dispersions.

\section{Conclusions}
This paper derives a non-Gaussian chance-constraint approach for stochastic nonlinear spacecraft targeting problems. The method employs a moment-informed ``banana" contour, using covariance for scale, and skew and kurtosis to capture the dominant bending and asymmetry of propagated uncertainty distributions. This enables direct geometric enforcement of half-plane constraints, including active-angle and smooth surrogate treatments of the resulting non-convex support problem. The ideas were demonstrated in two different numerical examples. The first was a simple asteroid orbiter control scenario, reproduced from prior literature. The second, more complex example revisited Artemis II stochastic guidance under the influence of navigation and maneuver exectution errors and circular restricted three-body dynamics. In both tests, the banana method substantially improved Monte Carlo constraint satisfaction over a linear covariance (LinCov) approach, while remaining computationally tractable at roughly an order-of-magnitude longer runtime than LinCov. This is a useful middle ground between LinCov and Monte Carlo methods in stochastic maneuver design, particularly for runtime-constrained applications, such as rapid trade studies or onboard guidance. This should be used when non-Gaussian distributions challenge the validity of a classical LinCov approach, but the degree of their non-Gaussianity is not extreme (i.e. the distribution is neither multi-modal nor exhibiting complex unmodeled geometry). 

\newpage
\appendix
\section{Appendix A: Conjugate Unscented Transform}
\setcounter{equation}{0}
\renewcommand{\theequation}{A.\arabic{equation}}
The equations below summarize CUT4,\cite{CUT_ACC} which estimates up to 4\textsuperscript{th}-order moments given an $N$-dimensional state with mean $\bm{\mu}$ and covariance $P$, subject to some (assumed nonlinear) process $\bm{g}$:
\begin{subequations}
\label{eq:cut4}
\begin{equation}
P=\bm{S}\bm{S}^\top,\qquad
\bm{\mathcal X}_i=\bm{\mu}+\bm{S}\bm{z}_i,\quad i=0,\dots,K-1, \qquad K=1+2N + 2^{N}.
\end{equation}
\begin{equation}
\bm{z}_0=\bm{0},\qquad
w_0=1-2Nw_1-2^{N}w_2 \ \ \ (\text{can freely choose } w_{0}=0 \ \text{if }N>2)
\end{equation}
\begin{equation}
\bm{z}_{k}^{\pm}=\pm r_1\,\bm{e}_k,\quad k=1,\dots,N,\qquad
(\text{uses weight } w_1)
\end{equation}
\begin{equation}
\tilde{\bm{z}}_{k}^{\pm}=\pm r_2\,\bm{\zeta}_{N,k},\quad
\|\bm{\zeta}_{N,k}\|=\sqrt{N},\quad k=1,\dots,2^{N},\qquad
(\text{uses weight } w_2)
\end{equation}
\begin{equation}
r_1=\sqrt{\frac{N+2}{2}},\qquad
r_2=\sqrt{\frac{N+2}{N-2}},\qquad
w_1=\frac{4}{(N+2)^2},\qquad
w_2=\frac{(N-2)^2}{2^{N}\,(N+2)^2}
\end{equation}
\begin{equation}
w_{i} = \bigg\{\begin{array}{cc} w_{1}, & 1 \leq i \leq 2N \\ w_{2}, & 2N + 1 \leq i\leq K \end{array}
\end{equation}
\begin{equation}
\bm{\mathcal Y}_i=\bm{g}(\bm{\mathcal X}_i),\qquad
\bm{\mu}_{\!y}=\sum_{i=1}^{K} w_i\,\bm{\mathcal Y}_i
\end{equation}
\begin{equation}
P_{\!y}=\sum_{i=1}^{K} w_i\big(\bm{\mathcal Y}_i-\bm{\mu}_{\!y}\big)\big(\bm{\mathcal Y}_i-\bm{\mu}_{\!y}\big)^\top
\end{equation}
\begin{equation}
M^{(3)}_{abc} = \sum_i w_i\,(\mathcal{Y}_{i,a}-\mu_a)(\mathcal{Y}_{i,b}-\mu_b)(\mathcal{Y}_{i,c}-\mu_c)
\end{equation}
\begin{equation}
M^{(4)}_{abcd} = \sum_i w_i\,(\mathcal{Y}_{i,a}-\mu_a)(\mathcal{Y}_{i,b}-\mu_b)(\mathcal{Y}_{i,c}-\mu_c)(\mathcal{Y}_{i,d}-\mu_d)
\end{equation}
\end{subequations}
Unlike an unscented transform (``UT"), which samples twice along each coordinate direction as $\pm\bm{e}_{k}$, CUT4 samples additionally along ``conjugate" axes $\bm{\zeta}_{N,k}$ which are weighted composite directions of the original basis. In general, for an $N$-dimensional
state, this includes the directions formed by assigning each coordinate either
``$+$'' or ``$-$'', resulting in $2^N$ extra off-axis samples in addition to the usual $2N$
axis-aligned ones. For example, in three dimensions these additional directions correspond to the eight
diagonals of a cube whose faces are normal to the three basis vectors and the six principal directions along these vectors. For $N=6$, to recover the first four statistical moments, CUT4 requires $2N + 2^{N}+1=77$ sigma points (or, by the free choice $w_{0}=0$, 76 points). CUT4 improves on the error properties of UT: For a weakly nonlinear transformation $\bm{g}(\bm{X})$, the expected covariance error is $\mathcal{O}(\|P\|^{5/2})$. For full details on the moment propagation schemes, please see References~\citenum{unscented,CUT_ACC}.
\section{Appendix B: Partial Derivatives for Chance Constraint Corrections}
\label{app:banana_support_partials}

For each control component, define
\begin{equation}
\label{eq:app_partial_operator}
\partial_j \equiv \frac{\partial}{\partial \chi_j},
\qquad j=1,2,3.
\end{equation}
The active support-gradient is
\begin{equation}
\label{eq:app_active_gradient}
\nabla_{\bm{\chi}}\psi(\bm{\chi},t^\star)
=
\nabla_{\bm{\chi}}A
+
\cos t^\star\nabla_{\bm{\chi}}B
+
\sin t^\star\nabla_{\bm{\chi}}C
+
\cos^2t^\star\nabla_{\bm{\chi}}D.
\end{equation}
The required Jacobian row is obtained from the componentwise partials of
$A,B,C,D$:
\begin{equation}
\label{eq:app_A_partial}
\partial_j A
=
\bm{n}^{\top}\partial_j\bm{\mu}
-
\left[
(\partial_jm_2)\alpha\sqrt{\lambda_2}
+
m_2(\partial_j\alpha)\sqrt{\lambda_2}
+
m_2\alpha\,\partial_j\sqrt{\lambda_2}
\right],
\end{equation}
\begin{equation}
\label{eq:app_B_partial}
\partial_j B
=
k
\left[
(\partial_jm_1)\sqrt{\lambda_1}
+
m_1\partial_j\sqrt{\lambda_1}
\right],
\end{equation}
\begin{equation}
\label{eq:app_C_partial}
\partial_j C
=
k
\left[
(\partial_jm_2)\sqrt{\lambda_2}
+
m_2\partial_j\sqrt{\lambda_2}
\right],
\end{equation}
and
\begin{equation}
\label{eq:app_D_partial}
\begin{aligned}
\partial_jD
={}&
(\partial_jm_1)c\sqrt{\lambda_1}
+
m_1(\partial_jc)\sqrt{\lambda_1}
+
m_1c\,\partial_j\sqrt{\lambda_1}
\\
&+
(\partial_jm_2)\alpha k^2\sqrt{\lambda_2}
+
m_2(\partial_j\alpha)k^2\sqrt{\lambda_2}
+
m_2\alpha k^2\partial_j\sqrt{\lambda_2}.
\end{aligned}
\end{equation}
Equations~\eqref{eq:app_A_partial}--\eqref{eq:app_D_partial} are the
components of $\nabla_{\bm{\chi}}A$, $\nabla_{\bm{\chi}}B$,
$\nabla_{\bm{\chi}}C$, and $\nabla_{\bm{\chi}}D$.

The constituent partials are as follows:
\begin{equation}
\label{eq:app_m_partial}
\partial_j\bm{m}
=
(\partial_jR)^{\top}\bm{n}.
\end{equation}
\begin{equation}
\label{eq:app_mu_partial}
\partial_j\bm{\mu}
=
S\bm{\Phi}_0B_{\chi,j}
+
S\partial_j\bm{\mu}_Y,
\end{equation}
where $B_{\chi,j}$ is column $j$ of
\begin{equation}
\label{eq:app_Bchi}
B_{\chi}
=
\frac{\partial \bar{\bm{X}}_0}{\partial \bm{\chi}}
=
\frac{1}{V}
\begin{bmatrix}
\bm{0}_{3\times3}\\
I_3
\end{bmatrix},
\qquad
V=\sqrt{\frac{\mu}{L}}.
\end{equation}

The eigenvalue partials associated with $\Sigma=R\Lambda R^\top$ are
\begin{equation}
\label{eq:app_lambda_partial}
\partial_j\lambda_i
=
\bm{e}_i^\top(\partial_j\Sigma)\bm{e}_i,
\qquad
\partial_j\sqrt{\lambda_i}
=
\frac{\partial_j\lambda_i}{2\sqrt{\lambda_i}}.
\end{equation}
For distinct eigenvalues,
\begin{equation}
\label{eq:app_R_partial}
\partial_j\bm{e}_i
=
\sum_{\ell\ne i}
\bm{e}_{\ell}
\frac{
\bm{e}_{\ell}^{\top}(\partial_j\Sigma)\bm{e}_i
}{
\lambda_i-\lambda_{\ell}
},
\qquad
\partial_jR=
\begin{bmatrix}
\partial_j\bm{e}_1 & \partial_j\bm{e}_2
\end{bmatrix}.
\end{equation}

In a code implementation one can, if desired, use a tuning parameter $\rho$ to interpolate between the theoretical LinCov-predicted covariance as the baseline, or the CUT-predicted. Using the latter produces a double-counting phenomenon which adds a natural ``buffer" to the banana prediction. While not completely rigorous, it could be a useful and natural tuning parameter in some circumstances. The covariance partial in this case is thus
\begin{equation}
\label{eq:app_sigma_partial}
\partial_j\Sigma
=
(1-\rho)\partial_j\Sigma_{\mathrm{lin}}
+
\rho\partial_j\Sigma_{\mathrm{CUT}},
\end{equation}
with
\begin{equation}
\label{eq:app_sigma_parts}
\partial_j\Sigma_{\mathrm{lin}}
=
S(\partial_jP_{\mathrm{lin}})S^\top,
\qquad
\partial_j\Sigma_{\mathrm{CUT}}
=
S(\partial_jP_Y)S^\top.
\end{equation}
The LinCov covariance partial is
\begin{equation}
\label{eq:app_Plin_partial}
\partial_jP_{\mathrm{lin}}
=
\bm{\Psi}_jP_0\bm{\Phi}_0^\top
+
\bm{\Phi}_0P_0\bm{\Psi}_j^\top,
\qquad
\bm{\Psi}_j=\partial_j\bm{\Phi}_0.
\end{equation}
In the implementation used here, $\rho=0$, so the covariance scaling of the
banana is LinCov-based, while the skewness and kurtosis corrections are
CUT-based.

For CUT point $i$,
\begin{equation}
\label{eq:app_Y_partial}
\partial_j\bm{Y}_i
=
(\bm{\Phi}_i-\bm{\Phi}_0)B_{\chi,j}.
\end{equation}
Thus, with CUT weights $w_i$,
\begin{equation}
\label{eq:app_muY_Z_partial}
\partial_j\bm{\mu}_Y
=
\sum_i w_i\partial_j\bm{Y}_i,
\qquad
\partial_j\bm{Z}_i
=
\partial_j\bm{Y}_i-\partial_j\bm{\mu}_Y.
\end{equation}
The CUT covariance and moment partials are
\begin{equation}
\label{eq:app_PY_partial}
\partial_jP_Y
=
\sum_i w_i
\left[
(\partial_j\bm{Z}_i)\bm{Z}_i^\top
+
\bm{Z}_i(\partial_j\bm{Z}_i)^\top
\right],
\end{equation}
\begin{equation}
\label{eq:app_M3_partial}
\partial_jM^{(3)}
=
\sum_i w_i
\left[
\partial_j\bm{Z}_i\otimes\bm{Z}_i\otimes\bm{Z}_i
+
\bm{Z}_i\otimes\partial_j\bm{Z}_i\otimes\bm{Z}_i
+
\bm{Z}_i\otimes\bm{Z}_i\otimes\partial_j\bm{Z}_i
\right],
\end{equation}
and
\begin{equation}
\label{eq:app_M4_partial}
\begin{aligned}
\partial_jM^{(4)}
=
\sum_i w_i
\big[
&
\partial_j\bm{Z}_i\otimes\bm{Z}_i\otimes\bm{Z}_i\otimes\bm{Z}_i
+
\bm{Z}_i\otimes\partial_j\bm{Z}_i\otimes\bm{Z}_i\otimes\bm{Z}_i
\\
&
+
\bm{Z}_i\otimes\bm{Z}_i\otimes\partial_j\bm{Z}_i\otimes\bm{Z}_i
+
\bm{Z}_i\otimes\bm{Z}_i\otimes\bm{Z}_i\otimes\partial_j\bm{Z}_i
\big].
\end{aligned}
\end{equation}

Let $W=\Lambda^{-1/2}R^\top$, with first and second rows $\bm{a}^\top$ and
$\bm{b}^\top$. Then
\begin{equation}
\label{eq:app_W_partial}
\partial_jW
=
\partial_j(\Lambda^{-1/2})R^\top
+
\Lambda^{-1/2}(\partial_jR)^\top,
\end{equation}
which gives $\partial_j\bm{a}$ and $\partial_j\bm{b}$.

Using the contraction notation e.g. 
\begin{equation}
\label{eq:app_contraction}
M^{(3)}[\bm{p},\bm{q},\bm{r}]
=
\sum_{a,b,c}p_aq_br_cM^{(3)}_{abc},
\end{equation}
the whitened-moment partials are
\begin{equation}
\label{eq:app_Euuu_partial}
\partial_jE_{uuu}
=
3M^{(3)}[\partial_j\bm{a},\bm{a},\bm{a}]
+
(\partial_jM^{(3)})[\bm{a},\bm{a},\bm{a}],
\end{equation}
\begin{equation}
\label{eq:app_Euuv_partial}
\partial_jE_{uuv}
=
M^{(3)}[\partial_j\bm{b},\bm{a},\bm{a}]
+
2M^{(3)}[\bm{b},\partial_j\bm{a},\bm{a}]
+
(\partial_jM^{(3)})[\bm{b},\bm{a},\bm{a}],
\end{equation}
and
\begin{equation}
\label{eq:app_Euuuu_partial}
\partial_jE_{uuuu}
=
4M^{(4)}[\partial_j\bm{a},\bm{a},\bm{a},\bm{a}]
+
(\partial_jM^{(4)})[\bm{a},\bm{a},\bm{a},\bm{a}].
\end{equation}

Finally, the two banana-correction partials appearing in
Eqs.~\eqref{eq:app_A_partial} and \eqref{eq:app_D_partial} are
\begin{equation}
\label{eq:app_c_partial}
\partial_jc
=
\frac{k^2-1}{6}\partial_jE_{uuu},
\end{equation}
and
\begin{equation}
\label{eq:app_alpha_partial}
\partial_j\alpha
=
\frac{
(\partial_jE_{uuv})(E_{uuuu}-1)
-
E_{uuv}(\partial_jE_{uuuu})
}{
(E_{uuuu}-1)^2
}.
\end{equation}
Substituting Eqs.~\eqref{eq:app_A_partial}--\eqref{eq:app_D_partial} into
Eq.~\eqref{eq:app_active_gradient} gives the Jacobian row associated with one
half-plane support constraint.

\section{Appendix C: Notes on Conservative Surrogate}
\label{app:log_integral_bound}

For a maximizer \(t^\star \in [0,2\pi]\), $\psi(t^\star)=M$. e make no requirement that there is a single unique maximizing angle $t^{\star}$, only that a maximum value $M$ exists. By the derivative bound \eqref{eq:def_L}, the following holds due to a Lipschitz condition:
\begin{equation}
\psi(t)\ge M - L|t-t^\star|,
\qquad t\in[0,2\pi].
\label{eq:app_lipschitz_cone}
\end{equation}
Exponentiation and integration of \eqref{eq:app_lipschitz_cone} gives
\begin{equation}
\frac{1}{2\pi}\int_{0}^{2\pi} e^{\psi(t)/\tau}\,dt
\ge
\frac{e^{M/\tau}}{2\pi}
\int_{0}^{2\pi} e^{-L|t-t^\star|/\tau}\,dt.
\label{eq:app_integral_lower_bound_step}
\end{equation}

Define the useful quantity $I(t^\star):=\int_{0}^{2\pi} e^{-L|t-t^\star|/\tau}\,dt$ and integrate:
\begin{equation}
I(t^\star)
=
\frac{\tau}{L}
\left(
2 - e^{-Lt^\star/\tau} - e^{-L(2\pi-t^\star)/\tau}
\right).
\label{eq:app_I_exact}
\end{equation}
Eq.~\eqref{eq:app_I_exact} is minimized when \(t^\star\) is at an endpoint, namely \(t^\star=0\) or \(t^\star=2\pi\). Consequently,
\begin{equation}
I(t^\star)
\ge
\frac{\tau}{L}\left(1-e^{-2\pi L/\tau}\right).
\label{eq:app_I_worst_case}
\end{equation}
Substituting \eqref{eq:app_I_worst_case} into \eqref{eq:app_integral_lower_bound_step} gives
\begin{equation}
\frac{1}{2\pi}\int_{0}^{2\pi} e^{\psi(t)/\tau}\,dt
\ge
e^{M/\tau}\,
\frac{\tau}{2\pi L}
\left(1-e^{-2\pi L/\tau}\right).
\label{eq:app_avg_exp_lower_bound}
\end{equation}
Taking \( \tau\log(\cdot) \) of both sides of \eqref{eq:app_avg_exp_lower_bound}
and rearranging gives the conservative upper bound
\begin{equation}
M
\le
g_{\text{b},\tau}[\psi]
-
\tau\log\!\left(
\frac{\tau}{2\pi L}\left(1-e^{-2\pi L/\tau}\right)
\right).
\label{eq:app_M_upper_bound}
\end{equation}
Using the definition of $\mathcal{C}(\tau,L)$ in Eq.~\eqref{eq:def_C}, Eq.~\eqref{eq:app_M_upper_bound} becomes
\begin{equation}
\max_t\psi(t)
\le
g_{\text{b},\tau}[\psi] + \mathcal{C}(\tau,L).
\label{eq:app_final_upper_bound}
\end{equation}

We must first note that the lack of smooth differentiability of Eq.~\eqref{eq:support_constraint_halfplane} is traded for that of Eq.~\eqref{eq:def_L}. However, this new discontinuity is much less severe, which we now show. Noting that in practice a very small $\tau$ should be chosen to minimize conservatism, we examine the behavior of Eq.~\eqref{eq:def_C} as \(\tau \to 0^+\):
\begin{equation}
\mathcal{C}(\tau,L)
\sim
\tau \log\!\left(\frac{2\pi L}{\tau}\right),
\qquad
\tau \to 0^+.
\label{eq:app_C_asymptotic}
\end{equation}

This observation is significant for two reasons. First, there is extremely weak dependence on $L$, so it can be over-estimated conservatively without ill effect. To see this, let \(L_2 = \rho L_1\) for some factor \(\rho\gg 1\). Then Eq.~\eqref{eq:app_C_asymptotic} gives
\begin{equation}
\mathcal{C}(\tau,L_2) - \mathcal{C}(\tau,L_1)
\sim
\tau \log\!\left(\frac{2\pi \rho L_1}{\tau}\right)
-
\tau \log\!\left(\frac{2\pi L_1}{\tau}\right)
=
\tau \log \rho.
\label{eq:app_C_difference}
\end{equation}
Hence even large changes in \(L\) modify $\mathcal{C}$ only by an amount proportional to \(\tau\), where $\tau\ll 1$. Numerical experiments show that even multiple order-of-magnitude overestimates of $L$ are inconsequential to the value of $\mathcal{C}$ when $\tau$ is chosen sufficiently small.

Second, differentiating Eq.~\eqref{eq:app_C_asymptotic} with respect to \(L\) gives
\begin{equation}
\frac{\partial \mathcal{C}}{\partial L}
\sim
\frac{\tau}{L},
\qquad
\tau \to 0^+.
\label{eq:app_dCdL_asymptotic}
\end{equation}
We remind the reader that $\tau$ is chosen as small as possible, and $L$ may be freely replaced with a highly conservative overestimate. Given the relative unimportance of $L$ for small $\tau$, we adopt a practice whereby $L$ may be treated as a \textit{fixed} or sequentially updated conservative bound, in which case the resulting surrogate is smooth in $\bm{\chi}$ for the purposes of the corrective step. In other words:
\begin{equation}
\nabla_{\bm{\chi}}g_{\text{a},\tau}[\psi] \approx \nabla_{\bm{\chi}}g_{\text{b},\tau}[\psi],
\qquad
\tau \to 0^+.
\label{eq:app_grad_surrogate_approx}
\end{equation}

To compute the gradients, the partial derivative of the log-integral term follows as:
\begin{equation}
\partial_j g_{\mathrm{b},\tau}[\psi]
=
\frac{
\int_0^{2\pi} e^{\psi(t)/\tau}\,\partial_j\psi(t)\,dt
}{
\int_0^{2\pi} e^{\psi(t)/\tau}\,dt
}.
\label{eq:app_gb_partial}
\end{equation}
Here
\begin{equation}
\partial_j\psi(\bm{\chi},t)
=
\partial_jA
+
\cos t\,\partial_jB
+
\sin t\,\partial_jC
+
\cos^2t\,\partial_jD,
\label{eq:app_psi_partial_integral}
\end{equation}
using the coefficient partials collected in Appendix B. In our implementation, $L$ is treated as a fixed or sequentially updated conservative
bound, so $\partial_j\mathcal{C}(\tau,L)$ is neglected,
$\partial_j g_{\mathrm{a},\tau}\approx \partial_j g_{\mathrm{b},\tau}$.
\bibliographystyle{AAS_publication}   
\bibliography{references}   

\end{document}